\documentclass[12pt]{article}
\usepackage{amsmath,amssymb}
\usepackage{url}
\usepackage{graphicx}
\makeatletter
\newif\if@preliminary
\@preliminaryfalse
\def\preliminary{\@preliminaryfalse}

\def\preprintno#1{\def\@preprintno{#1}}
\def\address#1{\def\@address{#1}}
\def\email#1#2{\thanks{\tt #1@{}#2}}
\def\abstract#1{\def\@abstract{#1}}
\renewcommand\abstractname{ABSTRACT}
\newlength\preprintnoskip
\newlength\abstractwidth
\@titlepagetrue
\renewcommand\maketitle{\begin{titlepage}%
  \let\footnotesize\small
  \hfill\parbox{\preprintnoskip}{%
  \begin{flushright}\@preprintno\end{flushright}}\hspace*{1cm}
  \vskip 60\p@
  \begin{center}%
    {\Large\bf\boldmath \@title \par}\vskip 1cm%
    {\sc\@author \par}\vskip 3mm%
    {\@address \par}%
    \if@preliminary
      \vskip 2cm {\large\sf NEXT-TO-FINAL DRAFT \par \@date}%
    \fi
  \end{center}\par
  \@thanks
  \vfill
  \begin{center}%
    \parbox{\abstractwidth}{\centerline{\abstractname}%
    \vskip 3mm%
    \@abstract}
  \end{center}
  \end{titlepage}%
  \setcounter{footnote}{0}%
  \let\thanks\relax\let\maketitle\relax
  \gdef\@thanks{}\gdef\@author{}\gdef\@address{}%
  \gdef\@title{}\gdef\@abstract{}\gdef\@preprintno{}
}%
\def\@citex[#1]#2{\if@filesw\immediate\write\@auxout{\string\citation{#2}}\fi
  \def\@citea{}\@cite{\@for\@citeb:=#2\do
    {\@citea\def\@citea{,\penalty\@m}\@ifundefined
       {b@\@citeb}{{\bf ?}\@warning
       {Citation `\@citeb' on page \thepage \space undefined}}%
\hbox{\csname b@\@citeb\endcsname}}}{#1}}
\def\citerange{\@ifnextchar [{\@tempswatrue\@citexr}{\@tempswafalse\@citexr[]}}
\def\@citexr[#1]#2{\if@filesw\immediate\write\@auxout{\string\citation{#2}}\fi
  \def\@citea{}\@cite{\@for\@citeb:=#2\do
    {\@citea\def\@citea{--\penalty\@m}\@ifundefined
       {b@\@citeb}{{\bf ?}\@warning
       {Citation `\@citeb' on page \thepage \space undefined}}%
\hbox{\csname b@\@citeb\endcsname}}}{#1}}
\long\def\@makecaption#1#2{%
  \vskip\abovecaptionskip
  \sbox\@tempboxa{#1: \emph{#2}}%
  \ifdim \wd\@tempboxa >\hsize
    #1: \emph{#2}\par
  \else
    \hbox to\hsize{\hfil\box\@tempboxa\hfil}%
  \fi
  \vskip\belowcaptionskip}
\def\fmslash{\@ifnextchar[{\fmsl@sh}{\fmsl@sh[0mu]}}
\def\fmsl@sh[#1]#2{%
  \mathchoice
    {\@fmsl@sh\displaystyle{#1}{#2}}%
    {\@fmsl@sh\textstyle{#1}{#2}}%
    {\@fmsl@sh\scriptstyle{#1}{#2}}%
    {\@fmsl@sh\scriptscriptstyle{#1}{#2}}}
\def\@fmsl@sh#1#2#3{\m@th\ooalign{$\hfil#1\mkern#2/\hfil$\crcr$#1#3$}}
\makeatother

\newcommand{\met}{p_T\hspace{-1.00em}/\hspace{0.4em}}
\newcommand{\me}{E\hspace{-0.6em}/\hspace{0.2em}}
\newcommand{\ie}{{\sl i.e.\ }}
\newcommand{\eg}{{\sl e.g.\ }}

\newcommand{\drbar}{\overline{\rm DR}}

\newcommand{\GeV}{{\ensuremath\rm GeV}}
\newcommand{\TeV}{{\ensuremath\rm TeV}}

\newcommand{\fb}{{\ensuremath\rm fb}}
\newcommand{\ab}{{\ensuremath\rm ab}}
\newcommand{\pb}{{\ensuremath\rm pb}}
\newcommand{\sel}{\tilde{e}_L}
\newcommand{\ser}{\tilde{e}_R}
\newcommand{\sml}{\tilde{\mu}_L}
\newcommand{\smr}{\tilde{\mu}_R}
\newcommand{\staul}{\tilde{\tau}_1}
\newcommand{\staur}{\tilde{\tau}_2}
\newcommand{\sne}{\tilde{\nu}_e}
\newcommand{\snm}{\tilde{\nu}_\mu}
\newcommand{\snt}{\tilde{\nu}_\tau}
\newcommand{\sul}{\tilde{u}_L}
\newcommand{\sur}{\tilde{u}_R}
\newcommand{\scl}{\tilde{c}_L}
\newcommand{\scr}{\tilde{c}_R}
\newcommand{\stl}{\tilde{t}_1}
\newcommand{\str}{\tilde{t}_2}
\newcommand{\sdl}{\tilde{d}_L}
\newcommand{\sdr}{\tilde{d}_R}
\newcommand{\ssl}{\tilde{s}_L}
\newcommand{\ssr}{\tilde{s}_R}
\newcommand{\sbl}{\tilde{b}_1}
\newcommand{\sbr}{\tilde{b}_2}
\newcommand{\neua}{\tilde{\chi}^0_1}
\newcommand{\neub}{\tilde{\chi}^0_2}
\newcommand{\neuc}{\tilde{\chi}^0_3}
\newcommand{\neud}{\tilde{\chi}^0_4}
\newcommand{\chap}{\tilde{\chi}_1^+}
\newcommand{\cham}{\tilde{\chi}_1^-}
\newcommand{\chbp}{\tilde{\chi}_2^+}
\newcommand{\chbm}{\tilde{\chi}_2^-}
\newcommand{\glo}{\tilde{g}}

\newcommand{\pone}{\phantom{0}}
\newcommand{\pten}{\phantom{00}}
\newcommand{\phun}{\phantom{000}}
\newcommand{\pth}{\phantom{0000}}
\newcommand{\tabc}[7]{$ #1 $ & #2 & #3 & #4 & #5 & #6 & #7 \\}
\newcommand{\mcent}{\multicolumn{1}{c}{---}}

\begin{document}
\preliminary        

\preprintno{DESY 05-068\\KEK-TH-1052\\MPP-2005-153\\WUE--ITP--2005--017\\[0.5\baselineskip]}

\title{Supersymmetry Simulations with Off-Shell Effects\\ for LHC and ILC}

\author{
 K.~Hagiwara\email{kaoru.hagiwara}{kek.jp}$^a$,
 W.~Kilian\email{wolfgang.kilian}{desy.de}$^b$,
 F.~Krauss\email{krauss}{theory.phy.tu-dresden.de}$^c$,
 T.~Ohl\email{ohl}{physik.uni-wuerzburg.de}$^d$,
 T.~Plehn\email{tilman.plehn}{cern.ch}$^e$,
 D.~Rainwater\email{rain}{pas.rochester.edu}$^f$,
 J.~Reuter\email{juergen.reuter}{desy.de}$^b$, 
 S.~Schumann\email{steffen}{theory.phy.tu-dresden.de}$^c$
}

\address{\it
$^a$Theory Division, KEK, Tsukuba, Japan\\
$^b$Theory Group, DESY, Hamburg, Germany\\
$^c$Institute for Theoretical Physics, University of Dresden, Germany\\
$^d$Institut f\"ur Theoretische Physik und Astrophysik, Universit\"at
  W\"urzburg, Germany\\
$^e$Heisenberg Fellow, Max Planck Institute for Physics,
  Munich, Germany\\
$^f$Dept. of Physics and Astronomy, University of Rochester,
  Rochester, NY, USA
\\[.5\baselineskip]
}

\abstract{
At the LHC and at an ILC, serious studies of new physics benefit from
a proper simulation of signals and backgrounds.  Using supersymmetric
sbottom pair production as an example, we show how multi-particle
final states are necessary to properly describe off-shell effects
induced by QCD, photon radiation, or by intermediate on-shell
states. To ensure the correctness of our findings we compare in detail
the implementation of the supersymmetric Lagrangian in {\sc madgraph},
{\sc sherpa} and {\sc whizard}.  As a future reference we give the
numerical results for several hundred cross sections for the
production of supersymmetric particles, checked with all three codes.
}

\maketitle


\section{Introduction}

The discoveries of the electroweak gauge bosons and the top quark more
than a decade ago established perturbative quantum field theory as a
common description of electromagnetic, weak, and strong interactions,
universally applicable for energies above the hadronic $\GeV$ scale.
The subsequent measurements of QCD and electroweak observables in
high-energy collision experiments at the SLAC SLC, CERN LEP, and
Fermilab Tevatron have validated this framework to an unprecedented
precision.  Nevertheless, the underlying mechanism of electroweak
symmetry breaking remains undetermined.  It is not clear how the
theory should be extrapolated beyond the electroweak scale
$v=246\;\GeV$ to the $\TeV$ scale or even higher energies~\cite{ewsb}.

\smallskip

At the LHC (and an ILC) this energy range will be directly probed for
the first time.  If the perturbative paradigm holds, we expect to see
fundamental scalar Higgs particles, as predicted by the Standard Model
(SM).  Weak-scale supersymmetry (SUSY) is a leading possible solution
to theoretical problems in electroweak symmetry breaking, and predicts
many additional new states.  The minimal supersymmetric extension of
the Standard Model (MSSM) is a model of softly-broken SUSY. The
supersymmetric particles (squarks, sleptons, charginos, neutralinos
and the gluino) can be massive in comparison to their SM counterparts.
Previous and current high-energy physics experiments have put
stringent lower bounds on supersymmetric particle masses, while
fine-tuning arguments lead us to believe they do not exceed a few
$\TeV$.  Therefore, a discovery in Run~II at the Tevatron is not
unlikely~\cite{tevatron}, and it will fall to the LHC to perform a
conclusive search for SUSY, starting in 2008.  Combining the energy
reach of the LHC with precision measurements at a possible future
electron-positron collider ILC, a thorough quantitative understanding
of the SUSY particles and interactions would be
possible~\cite{lhc-ilc}.

Most realistically, SUSY will give us a plethora of particle
production and decay channels that need to be disentangled and
separated from the SM background.  To uncover the nature of
electroweak symmetry breaking we not only have to experimentally
analyze multi-particle production and decay signatures, we also need
to accurately simulate the model predictions on the theory side.

\medskip

Much SUSY phenomenology has been performed over the years in
preparation for LHC and ILC, nearly all of it based on relatively
simple $2\to 2$ processes~\cite{lhc_tree,ilc_tree} or their
next-to-leading order (NLO) corrections.  These approximations are
useful for highly inclusive analyses and convenient for analytical
calculations, but should be dropped once we are interested in precise
measurements and their theoretical understanding.  Furthermore, for a
proper description of data, we need Monte-Carlo event generators that
fully account for high-energy collider environments.  Examples for
necessary improvements include: consideration of spin
correlations~\cite{susy_spins} and finite width effects in
supersymmetric particle decays~\cite{cascade}; SUSY-electroweak and
Yukawa interferences to some SUSY-QCD processes; exact rather than
common virtual squark masses; and $2\to 3$ or $2\to 4$ particle
production processes such as the production of hard jets in SUSY-QCD
processes~\cite{skands} or SUSY particles produced in weak-boson
fusion (WBF)~\cite{WBF}.

\medskip

In this paper we present three new next-generation event
generators for SUSY processes: {\sc madgraph ii}/{\sc
madevent}~\cite{madgraph,madevent}, {\sc o'mega}/{\sc
whizard}~\cite{Omega,Whizard}, and {\sc amegic}\texttt{++}/{\sc
sherpa}~\cite{amegic,sherpa}.  They properly take into account various
physics aspects which are usually approximated in the literature, such
as those listed above.  They build upon new methods and algorithms for
automatic tree-level matrix element calculation and phase space
generation that have successfully been applied to SM
phenomenology~\cite{SM-MC,comphep,grace}.  Adapted to the more
involved structure of the MSSM, they are powerful tools for a new
round of MSSM phenomenology, especially at hadron colliders.

\medskip

The structure of the paper is as follows: In Sec.~\ref{sec:susysim} we
consider basic requirements for realistic SUSY simulations, in
particular the setup of consistent calculational rules and conventions
as a \emph{conditio sine qua non} for obtaining correct and
reproducible results.  Sec.~\ref{sec:programs} gives some details of
the implementation of MSSM multi-particle processes in the three
generators, while Sec.~\ref{sec:comparison} is devoted to numerical
checks.  Finally, Secs.~\ref{sec:lhc} and~\ref{sec:ilc} cover one
particular application, the physics of sbottom squarks at the LHC and
an ILC, respectively.  Our emphasis lies on off-shell effects of
various kinds which for the first time we accurately describe using
the tools presented in this paper.

In the extensive Appendix we list as a future reference cross sections
for several hundred SUSY $2\to 2$ processes that are the main part of
these checks.  We include all information necessary to reproduce these
numbers.


\section{Supersymmetry Simulations}
\label{sec:susysim}

Throughout this paper, we assume R-parity conservation in the MSSM.
The SUSY particle content consists of the SM particles, the five Higgs
bosons, and their superpartners, namely six sleptons, three
sneutrinos, six up-type and six down-type squarks, two charginos, four
neutralinos, and the gluino.  We allow for a general set of TeV-scale
(or weak-scale) MSSM parameters, with a few simplifying restrictions:
(i) we assume CP conservation, \ie all soft-breaking terms in the
Lagrangian are real (cf.~also (iii) below); (ii) we neglect masses and
Yukawa couplings for the first two fermion generations, \ie left-right
mixing occurs only for third-generation squarks and sleptons; (iii)
correspondingly, we assume the SM flavor structure to be trivial,
$V_{\rm CKM}=V_{\rm MNS}=1$; (iv) we likewise assume the flavor
structure of SUSY-breaking terms to be trivial.

None of these simplifications is a technical requirement, and all
codes are capable of dealing with complex couplings as well as
arbitrary fermion and sfermion mass and mixing matrices.  However,
with very few exceptions, these effects are numerically unimportant or
irrelevant for the simulation of SUSY scattering and decay processes
at high-energy colliders.  (In Sec.~\ref{sec:ckm} we discuss residual
effects of nontrivial flavor structure.)

We thus define the MSSM as the general TeV/weak-scale Lagrangian for
the SM particles with two Higgs doublets, with gauge- and
Lorentz-invariant, R-parity-conserving, renormalizable couplings, and
softly-broken supersymmetry.  Unfortunately, while this completely
fixes the physics, it leaves a considerable freedom in choosing phase
conventions.  The large number of Lagrangian terms leaves ample room
for error in deriving Feynman rules, coding them in a computer
program, and relating the input parameters to a standard convention.

\medskip

The three codes we consider here are completely independent in their
derivation of Feynman rules, implementation, matrix element
generation, phase space setup, and integration methods.  A detailed
numerical comparison should therefore reveal any mistake in these
steps.  To this end, we list a set of $2\to 2$ scattering processes
that involve all Feynman rules that could be of any relevance in
Sec.~\ref{sec:comparison}.


\subsection{Parameters and Conventions}

Apart from the simplifications listed above, we do not make any
assumptions about SUSY breaking.  No physical parameters are
hard-coded into the programs.  Instead, all codes use a set of
weak-scale parameters in the form of a SUSY Les Houches Accord (SLHA)
input file~\cite{SLHA}.  This file may be generated by any one of the
standard SUSY spectrum generators~\cite{Softsusy,SUSY-spectrum}.

Since the SLHA defines weak-scale parameters in a particular
renormalization scheme, we have to specify how to use them for our
tree-level calculations: we fix the electroweak parameters via $G_F$,
$M_Z$, and $\alpha_{\rm QED}$.  Using the tree-level relations (as
required for gauge-invariant matrix elements at tree level) we obtain
parameters such as $\sin^2\theta_w$ and $M_W$ as derived quantities;
$M_W$ and $M_Z$ are defined as pole masses.

The SLHA uses pole masses for all MSSM particles, while mixing
matrices and Yukawa couplings are given as loop-improved $\drbar$
values.  From this input we need to derive a set of mass and coupling
parameters suitable for comparing tree-level matrix-element
calculations.  This leads to violation of electroweak gauge invariance
which we discuss in Sec.~\ref{sec:unitarity}.  However, numerically
this is a minor problem, relevant only for some processes (\eg SUSY
particle production in weak-boson fusion~\cite{WBF}) at asymptotically
high energies.  For the numerical results of this paper, we therefore
use the SLHA masses and mixing matrices at face value.

For the bottom and top quarks we identify the (running) Yukawa
couplings and the masses, as required by gauge invariance.  The weak
scale as the renormalization point yields realistic values for the
Yukawa couplings.  One might be concerned that the kinematical masses
are then off from their actual values.  However, since our production
cross section should be regarded as the leading contribution to the
inclusive cross section, the relevant scale is the energy scale of the
whole process rather than the scale of individual heavy quarks.  This
necessitates the use of running masses to make a reliable estimate.

The trilinear couplings and the $\mu$ parameter which explicitly
appear in some couplings are fixed by the off-diagonal entries in the
chargino, neutralino and sfermion mass matrices. We adopt two schemes
for negative neutralino mass eigenvalues: {\sc sherpa} and {\sc
madgraph} use the negative values directly in the propagator and wave
function.  {\sc o'mega}/{\sc whizard} rotates the neutralino fields to
positive masses, which yields a complex neutralino mixing matrix, even
though CP is conserved.

For our comparison we neglect all couplings that contain masses of
light-flavor fermions, \ie the Higgs couplings to first- and
second-generation fermions and their supersymmetric counterparts; as
well as left--right sfermion mixing. This includes neglecting light
fermion masses in the neutralino and chargino sector, which would
otherwise appear via Yukawa-higgsino couplings.  Physically, this is
motivated by flavor constraints which forbid large deviations from
universality in the first and second generations~\cite{flavor_review}.
For our LHC calculations we employ CTEQ5 parton distribution 
functions~\cite{Lai:1999wy}.


\subsection{Unitarity and the SLHA Convention}
\label{sec:unitarity}

The MSSM is a renormalizable quantum field theory~\cite{dominik}.  To
any fixed order in perturbation theory, a partial-wave amplitude
calculated from the Feynman rules, renormalized properly, is bounded
from above.  Cross sections with a finite number of partial waves (\eg
$s$-channel processes) asymptotically fall off like $1/s$, while
massless particle exchange must not lead to more than a logarithmic
increase with energy.  This makes unitarity a convenient check for the
Feynman rules in our matrix element calculators.

As an example, individual diagrams that contribute to $2\to 2$ weak
boson scattering rise like the fourth power of the energy, but the two
leading terms of the energy expansion cancel among diagrams to
ameliorate this to a constant.  This property connects the three- and
four-boson vertices, and predicts the existence and couplings of a
Higgs boson, assuming the theory is weakly interacting to high
energies~\cite{Uni}.  For example, for weak boson fusion to
neutralinos and charginos, these unitarity cancellations can be neatly
summarized in a set of sum rules for the SUSY masses and
couplings~\cite{WBF}.  For generic Higgs sectors, the unitarity
relations were worked out in~\cite{Gunion:1990kf}.

Many, but not all, terms in the Lagrangian can be checked by requiring
unitarity.  For instance, gauge cancellations in $WW$ scattering to
two SUSY particles need not happen if the final-state particle has an
$SU(2)\times U(1)$ invariant mass term.  In the softly-broken SUSY
Lagrangian, this property holds for the gauginos and higgsinos as well
as for the second Higgs doublet in the MSSM.  For these particles, we
expect unitarity relations to impose some restrictions on their
couplings, but not a complete set of equations, so some couplings
remain unconstrained.

\medskip

As mentioned above, for our numerical comparison of SUSY processes we
use a renormaliza\-tion-group improved spectrum in the SLHA
format~\cite{Softsusy,SUSY-spectrum}.  In particular, we adopt this
spectrum for the Higgs sector, where gauge invariance (or unitarity)
relates masses, trilinear and quartic couplings.  While at tree-level
all unitarity relations are automatically satisfied, any improved
spectrum will violate unitarity constraints unless the Higgs trilinear
couplings are computed in the same scheme.  However, not all couplings
are known to the same accuracy as the Higgs
masses~\cite{higgs_twoloop}.  We follow the standard approach of
computing the trilinear Higgs couplings from effective mixing angles
$\alpha$ and $\beta$.  As a consequence, we expect unitarity
violation.  Luckily, this only occurs in $2 \to 3$ processes of the
type $WW\to WWH$~\cite{Gunion:1990kf}, while in $2\to 2$ processes of
the type $WW\to HH$ where one might naively expect unitarity
violation, the values of the Higgs trilinear couplings change the
value of total high-energy asymptotic cross section but do not affect
unitarity.

A similar problem arises in the neutralino and chargino sector.
Unitarity is violated at high energies in processes of the type
$VV\to\tilde\chi\tilde\chi$ ($V=W,Z$)~\cite{WBF}.  If we use the
renormalization-group improved $\drbar$ neutralino and chargino mass
matrices (or equivalently the masses and mixing matrices) the
gaugino--higgsino mixing entries which are equivalent to the Higgs
couplings of the neutralinos and the charginos implicitly involve
$M_{W,Z}$, also in the $\drbar$ scheme.  To ensure proper gauge
cancellations which guarantee unitarity, these gauge boson masses must
be identical to the kinematical masses of the gauge bosons in the
scattering process, which are usually defined in the on-shell scheme.
One possible solution would be to extract a set of gauge boson masses
that satisfies all tree-level relations from the mass matrices.  This
scheme has the disadvantage that while it works for the leading
corrections, it will likely not be possible to derive a consistent set
of weak parameters in general.  Moreover, the higher-order corrections
included in the renormalization-group improved neutralino and chargino
mass matrices will not be identical to the leading corrections to, for
example, the $s$-channel propagator mass.  However, an artificial
spectrum that is specifically designed to fulfill the tree-level
relations can be used for a technical test of high-energy unitarity.
Such a detailed check has been performed for the {\sc susy-madgraph}
implementation~\cite{WBF}.


\subsection{Symmetries and Ward Identities}

An independent method for verifying the implementation is the
numerical test of symmetries and their associated Ward identities.  A
trivial check is provided by the permutation and crossing symmetries
of many-particle amplitudes.  More subtle are the Ward identities of
gauge symmetries, which can be tested by replacing the polarization
vector of any one external gauge boson by its momentum
$\epsilon_\mu(k)\to k_\mu$ and, if necessary, subtracting the
amplitude with the corresponding Goldstone boson amplitude.  Finally,
the SUSY Ward identities can be tested numerically.

Ward identities have the advantage that they require no additional
computer program, can be constructed automatically and can be applied
separately for each point in phase space.  If applied in sensitive
regions of phase space, tests of Ward identities will reveal numerical
instabilities.  Extensive tests of this kind have been carried out for
the matrix elements generated by {\sc o'mega} and its associated
numerical library for the SM~\cite{Schwinn:2003fp} and for the
MSSM~\cite{Ohl:2002jp}.


\subsection{Intermediate Heavy States}
\label{sec:multi}

During the initial phase of the LHC, narrow resonances can be
described by simple $2\to2$ production cross sections and subsequent
cascade decays.  However, establishing that these resonances are
indeed the long-sought SUSY partners would call for more sophisticated
tools.

The identification of resonances as SUSY partners would require
determination of their spin and parity quantum
numbers~\cite{susy_spins}.  This in turn requires a proper description
of the spin correlations among the particles in the production and the
decay cascades.  The simplest consistent approximation calculates the
Feynman diagrams for the $2\to n$ process and forces narrow
intermediate states on the mass shell without affecting spin
correlations.  For fermions we can write the leading term in the small
expansion parameter $\Gamma/m$ as:
\begin{equation}\label{eq:on-shell}
  \frac{1}{|s-m+\mathrm{i}\Gamma|^2}
    \to \frac{\pi}{m\Gamma} \, \delta(s-m^2)
\end{equation}
For SM processes this computation of $2\to n$ matrix elements has been
successfully automatized by the programs described below.  The
alternative approach of manually inserting the appropriate density
matrices for production and decay is more error-prone due to the need
for consistent phase conventions.

\smallskip

The width of the heavy resonances are themselves observables predicted
by SUSY for a given set of soft breaking parameters and should be
taken into account.  A na\"ive Lorentzian smearing of
Eq.~(\ref{eq:on-shell}) will not yield a theoretically consistent
description of finite width effects.  Gauge and SUSY Ward identities
are immediately violated once amplitudes are continued off-shell.
Since scattering amplitudes in gauge theories and SUSY theories
exhibit strong numerical cancellations, the violation of the
corresponding Ward identities can result in numerically large effects.
Therefore a proper description of a resonance with a finite width
requires a complete gauge invariant set of diagrams, the simplest of
which is the set of all diagrams contributing to the $2\to n$
process~\cite{gauge_tree}.  In Secs.~\ref{sec:lhc} and~\ref{sec:ilc}
we study the numerical impact of finite-width effects for the concrete
example of sbottom production at high-energy colliders.

Intermediate charged particles with finite widths present additional
gauge invariance issues, which were studied at LEP2 in great detail
for $W$ boson production processes~\cite{LEP2-Gauge-Gospel}.  Although
various prescriptions for widths are available in the matrix element
generators described in the paper, we used the fixed-width scheme for
the calculations.  A careful analysis on the impact of different
choices is beyond the scope of the paper.


\section{Calculational Methods and Algorithms}
\label{sec:programs}

Each of the three calculational tools we use for this paper consists
of two independent programs.  The first program uses a set of Feynman
rules, which can be preset or user-defined, to generate computer code
that numerically computes the tree-level scattering amplitude for a
chosen process.  These numerical codes call library functions to
compute wave functions of external particles, internal currents and
vertices to obtain the complete helicity amplitude (the amplitude for
all helicity configurations of external particles, which are then
summed over~\cite{helamp}).  The second program performs adaptive
phase space integration and event generation, and produces integrated
cross sections and weighted or unweighted event samples.  The required
phase space mappings are determined automatically, using appropriate
heuristics, from the `important' Feynman diagrams contributing to the
process that is being studied.

In principle, there is nothing that precludes the use of other
combinations of the three matrix element generators and the three
phase space integrators.  In practice however, it requires some effort
to adopt the interfaces that have grown organically.  Nevertheless,
{\sc whizard} can \eg use {\sc madgraph} as an alternative to {\sc
o'mega}.

{\sc helas}~\cite{HELAS} is the archetypal helicity amplitude library
and is now employed by many automated matrix element generators.  The
elimination of common subexpression to optimize the numerical
evaluation was already suggested in Refs.~\cite{HELAS} for the manual
construction of scattering amplitudes.  The actual libraries used by
our three tools choose between different trade-offs of
maintainability, extensibility, efficiency and numerical accuracy.

\medskip

Majorana spinors are the crucial new ingredient for calculating
helicity amplitudes in supersymmetric field theories.  In the simple
example process $e^+e^-\to\tilde\chi\tilde\chi$ we see the
complication which arises: if we na\"ively follow the fermion number
flow of the incoming fermions, the $t$-channel and $u$-channel
amplitudes require different external spinors for the final-state
fermions.  The most elegant algorithm known for unambiguously
assigning a fermion flow and the relative signs among Feynman diagrams
is described in Ref.~\cite{Denner:Majorana}.  Consequently, all matrix
element generators use an implementation of this algorithm.

\smallskip

Beneath some common general features, the similarities of the three
tools quickly disappear: they use different algorithms, implemented in
different programming languages.  That such vastly different programs
can be tested against each other with a Lagrangian as complex as that
of the TeV-scale MSSM should give confidence in the predictive power
of these programs for SUSY physics at the LHC and later at an ILC.

\medskip

To compute cross sections in the MSSM, we need a consistent set of
particle masses and mixing matrices, computed for a chosen
SUSY-breaking scenario.  Various spectrum generators are available,
all using the SUSY Les Houches Accord as their spectrum interface.
The partonic events generated by our three tools can either be
fragmented by a built-in algorithm ({\sc sherpa}) or passed via a
standard interface to external hadronization
packages~\cite{event_lha}.  However, proper hadronization of the
parton level results presented here is beyond the scope of this paper.


\subsection{{\sc madgraph ii} and {\sc madevent}}

{\sc madgraph}~\cite{madgraph} was the first program allowing fully
automated calculations of squared helicity amplitudes in the Standard
Model.  In addition to being applied to many physics calculations, it
was later frequently used as a benchmark for testing the accuracy of
new programs as well as for gauging the improvements implemented in
the new programs.

\medskip

{\sc madgraph ii} is implemented in {\sc fortran77}.  It generates all
Feynman diagrams for a given process, performs the color algebra and
translates the result into a {\sc fortran77} procedure with calls to
the {\sc helas} library.  During this translation, redundant
subexpressions are recognized and computed only once.  While the
complexity continues to grow asymptotically with the number of Feynman
diagrams, this approach generates efficient code for typical
applications.

The correct implementation of color flows for hadron collider physics
was an important objective for the very first version of {\sc
madgraph}, while the implementation of extensions of the standard
model remained nontrivial for users.  {\sc madgraph ii} reads the
model information from two files and supports Majorana fermions,
allowing fully automated calculations in the MSSM.  The MSSM
implementation makes use and extends the list of Feynman rules that
have been derived in the context of~\cite{prospino,higgs_pair,kaoru}.

\medskip

{\sc madevent}~\cite{madevent} uses phase space mappings based on
single squared Feynman diagrams for adaptive multi-channel
sampling~\cite{Kleiss:1994qy}.  The {\sc madgraph}/{\sc madevent}
package has a web-based user interface and supports shortcuts such as
summing over initial state partons, summing over jet flavors and
restricting intermediate states.  Interfaces to parton shower and
hadronization Monte Carlos~\cite{event_lha} are available.


\subsection{{\sc o'mega} and {\sc whizard}}

{\sc o'mega}~\cite{Omega} and {\sc whizard}~\cite{Whizard} were
initially designed for $e^+e^-$ linear colliders studies.  {\sc
o'mega} constructs numerically stable and optimally factorized
scattering amplitudes and allows the study of physics beyond the
Standard Model.  A general treatment of color was added to {\sc
o'mega} only recently and is currently available only in conjunction
with {\sc whizard}.

\medskip

{\sc o'mega} constructs an expression for the scattering matrix
element from a description of the Feynman rules and the target
programming language.  The complexity of these expressions grows only
exponentially with the number of external particles, unlike the
factorial growth of the number of Feynman diagrams.  Optionally, {\sc
o'mega} can calculate cascades: long-lived intermediate particles can
be forced on the mass shell in order to obtain gauge invariant
approximations with full spin correlations.

{\sc o'mega} is implemented in the functional programming language
Objective Caml~\cite{O'Caml}, but the compiler is portable and no
knowledge of Objective Caml is required for using {\sc o'mega} with
the supported models.  The tables describing the Lagrangians can be
extended by users. Its set of MSSM Feynman rules was derived in
accordance with Ref.~\cite{reu_kur}.

\medskip

{\sc whizard} builds a Monte Carlo event generator on the library
VAMP~\cite{vamp} for adaptive multi-channel sampling.  It uses
heuristics to construct phase space parameterizations corresponding to
the dominant time- and space-like singularities for each process.  For
processes with many identical particles in the final state, symmetries
are used extensively to reduce the number of independent channels.

{\sc whizard} is written in {\sc fortran95}, with some Perl glue code.
It is particularly easy to simulate multiple processes (\ie reducible
backgrounds) with the correct relative rates simultaneously.  It has
an integrated interface to {\sc pythia}~\cite{pythia} that follows the
Les Houches Accord~\cite{event_lha} for parton showers and
hadronization.


\subsection{{\sc amegic}\texttt{++} and {\sc sherpa}}

{\sc sherpa}~\cite{sherpa} is a new complete Monte Carlo Generator for
collider physics, including hard matrix elements, parton showers,
hadronization and other soft physics aspects, written from scratch in
\texttt{C++}.  The key feature of {\sc sherpa} is the implementation of 
an algorithm~\cite{Catani:2001cc,Krauss:2002up,Schalicke:2005nv},
which allows consistent combination of tree-level matrix elements for
the hard production of particles with the subsequent parton showers
that model softer bremsstrahlung.  This algorithm has been tested in
various processes~\cite{Krauss:2004bs,Gleisberg:2005qq}.  Both of the
other programs described above connect their results with full event
simulation through interfaces to external programs.

\medskip

{\sc amegic}\texttt{++}~\cite{amegic} is the matrix element generator
for {\sc sherpa}.  It generates all Feynman diagrams for a process
from a model description in \texttt{C++}.  Before writing the
numerical code to evaluate the complete amplitude (including color
flows), it eliminates common subexpressions.  {\it En passant}, it
selects appropriate phase space mappings for multi-channel sampling
and event generation~\cite{Kleiss:1994qy}.  For integration it relies
on {\sc vegas}~\cite{Lepage:1980dq} to further improve the efficiency
of the dominant integration channels.  The MSSM Feynman rules and
conventions in {\sc amegic}\texttt{++} are taken from
Ref.~\cite{rosiek}.


\section{Pair Production of SUSY Particles}
\label{sec:comparison}


\subsection{The Setup}
\label{sec:setup}

As long as R-parity is conserved, SUSY particles are only produced in
pairs.  Therefore, SUSY phenomenology at the LHC and ILC amounts to
essentially searching for all accessible supersymmetric
pair-production channels with subsequent (cascade) decays.  Proper
simulations need to describe this type of processes as accurately as
possible.  This requires a careful treatment of many-particle final
states, off-shell effects and SUSY as well as SM backgrounds.  The
complexity of this task and the variety of conventions and schemes
commonly used require careful cross-checks at all levels of the
calculation.

As a first step, we present a comprehensive list of total cross
sections for on-shell supersymmetric pair production processes (cf.\
Appendix~\ref{sec:xsec}).  These results give a rough overview of the
possible SUSY phenomenology at future colliders, at least for the
chosen point in SUSY parameter space.  The second purpose of this
computation is a careful check of our sets of Feynman rules and their
numerical implementation.  After testing our tools we will then move
on to a proper treatment beyond na\"ive $2 \to 2$ production
processes.  We compute all numbers independently with {\sc madgraph},
{\sc whizard}, and {\sc sherpa}, using identical input parameters.  We
adopt a MSSM parameter set that corresponds to the point
SPS1a~\cite{SPS1a}.  This point assumes gravity mediated supersymmetry
breaking with the universal GUT-scale parameters:
\begin{equation}\label{SPS1a}
  m_0 = 100      \;\GeV \; , \qquad
  m_{1/2} = 250  \;\GeV \; , \qquad
  A_0 = -100     \;\GeV \; , \qquad
  \tan\beta = 10 \;        , \qquad
  \mu > 0        \; .
\end{equation}
We use {\sc softsusy} to compute the TeV-scale physical
spectrum~\cite{Softsusy}.  For the purpose of evaluating $2\to 2$
cross sections, we set all SUSY particle widths to zero.  The final
states are all possible combinations of two SUSY partners or two Higgs
bosons.  The initial states required to test all the SUSY vertices
are:
\begin{gather*}
  e^+e^-,\ e^-\bar\nu_e,\ e^-e^-,\ \tau^+\tau^-,\ \tau^-\bar\nu_\tau,\
  u\bar{u},\ d\bar{d},\ uu,\ dd,\ b\bar{b},\ b\bar{t},\ \\
  W^+W^-,\ W^-Z,\ W^-\gamma,\ ZZ,\ Z\gamma,\ \gamma\gamma,\
  gW^-,\ gZ,\ g\gamma,\ gg,\ ug,\ dg \, .
\end{gather*}
The (partonic) initial-state energy is always fixed.  This allows for
a comparison of cross sections without dependence on parton structure
functions, and with much-improved numerical efficiency.  Clearly, some
of these initial states cannot be realized on-shell or are even
impossible to realize at a collider.  They serve only as tests of the
Feynman rules.  Any MSSM Feynman rule relevant for an observable
collider process is involved in at least one of the considered
processes.  For SM processes, comprehensive checks and comparisons
were performed in the past~\cite{lusifer}.

\medskip

The complete list of input parameters is given in
Appendix~\ref{sec:input}.  The input is specified in the SLHA
format~\cite{SLHA}.  This ensures compatibility of the input
conventions, even though different conventions for the Lagrangian and
Feynman rules are used by the different programs.

In Appendix~\ref{sec:xsec}, we list and compare the results for two
partonic c.m.\ energies $\sqrt{s}=500\;\GeV$ and $2\;\TeV$.  All
results agree within a Monte Carlo statistical uncertainty of $0.1\%$
or less.  These errors reflect neither the accuracy nor the efficiency
of any of the programs; we do not specify the number of matrix element
calls or the amount of CPU time required in the computation.  To
obtain a precise $2\to 2$ total cross section, Monte Carlo integration
is not a good choice.  On the other hand these simple processes serve
as the most efficient framework to test the numerical implementation
of Feynman rules and the MSSM spectrum.

We emphasize that the three programs {\sc madgraph}, {\sc whizard},
and {\sc sherpa}, and their SUSY implementation are completely
independent.  As explained in Sec.~\ref{sec:programs}, they use
different conventions, signs and phase choices for the MSSM Feynman
rules; have independent algorithms and {\sc helas}-type libraries; and
use different methods for parameterizing and sampling the phase space.
We consider our results a strong check that covers all practical
aspects of MSSM calculations, from the model setup to the numerical
details.  Specifically, we confirm the Feynman rules in
Ref.~\cite{rosiek} as they are used in {\sc sherpa}.  These Feynman
rules do not use the SLHA format, so translating them is a non-trivial
part of the implementation.  For {\sc madgraph} and {\sc whizard}, the
Feynman rules were derived independently.


\subsection{Sample Cross Sections}

We briefly discuss our cross section results and their physical
interpretation.  While the numbers are specific for the chosen point
SPS1a~\cite{SPS1a} and its associated mass spectrum, many features of
the results are rather generic in one-scale SUSY breaking models and
depend only on the structure of the TeV-scale MSSM.


\subsubsection{$\mathbf{e^+e^-}$ processes}

All $e^+e^-$--induced SUSY production cross sections receive
contributions from $s$-channel $Z$ and (for charged particles) photon
exchange.  The couplings of the supersymmetric particles to $Z$ and
photon are determined by the $SU(2)_L\times U(1)_Y$ gauge couplings
and mixing angle.  As expected from perturbative unitarity, all
$s$-channel-process cross sections asymptotically fall off like $1/s$.
If the process in question includes $t$-channel exchange, then we must
sum over all partial waves
\begin{equation}
\frac{m^2}{2p\cdot k + m^2} \; = \;
\sum^\infty_{n=0} \biggl( \frac{-2p\cdot k}{m^2} \biggr)^n
\Longrightarrow\quad
\sigma \propto \frac{1}{s} \times 
\begin{cases}
         \log \frac{s}{m^2} & \text{for no vector boson exchange,} \\
\phantom{\log}\frac{s}{m^2} & \text{for vector boson exchange.}
\end{cases}
\end{equation}
The implication of the second line is that Coulomb scattering, WBF,
and in some sense all hadronic cross sections, do not decrease with
$s$.

We show the $e^+e^-$ cross sections in Table~\ref{xsec:e+e-}.  The
largest, of up to a few hundred~fb at $\sqrt{s}=500\;\GeV$, correspond
to sneutrino and selectron production, $\neua$ and $\neub$, and
chargino pair production.  These are the processes with a dominant
$t$-channel slepton contribution.  In SPS1a the heavier neutralinos
$\neuc,\neud$ are almost pure higgsinos.  Higgsinos couple only to the
$s$-channel $Z$, and diagonal pair production of $\neuc\neuc,
\neud\neud$ is suppressed because of the inherent cancellation between
the two higgsino fractions $h_u$ and $h_d$; \ie the amplitudes are
proportional to $|h_u|^2-|h_d|^2$, which vanishes in the limit where
they have the same higgsino masses.  Only mixed $\neuc\neud$
production has a significant cross section, because it is proportional
to the sum $|h_u|^2+|h_d|^2$.

In the Higgs sector, SPS1a realizes the decoupling limit where the
light Higgs $h$ closely resembles the SM Higgs.  The production
channels $Zh$, $AH$, and $H^+H^-$ dominate if kinematically
accessible, while the reduced coupling of the $Z$ to heavy Higgses
strongly suppresses the $ZH$ and $Ah$ channels.

For completeness, we also show the $e^-\bar\nu_e$ set of cross
sections in Table~\ref{xsec:e-nu}, even though such a collider is
infeasible.


\subsubsection{$\mathbf{W^+W^-}$ and $\mathbf{WZ}$ processes}

The cross sections for weak boson fusion processes, shown in
Tables~\ref{xsec:W+W-} and~\ref{xsec:W-Z}, are generically of the same
order of magnitude as their fermion-initiated counterparts, with a few
notable differences.  In addition to gauge boson exchange, $s$- and
$t$-channel Higgs exchange contributes to WBF production of
third-generation sfermions, neutralinos, charginos, and Higgs/vector
bosons.  These processes are sensitive to a plethora of Higgs
couplings to supersymmetric particles.  Furthermore, the longitudinal
polarization components of the external vector bosons approximate, in
the high-energy limit, the pseudo-Goldstone bosons of electroweak
symmetry breaking.  This results in a characteristic asymptotic
behavior (that can be checked by inserting $\sqrt{s}$ values of
several $\TeV$, not shown in the tables): the total cross sections for
vector-boson and CP-even Higgs pair production in WBF approach a
constant at high energy, corresponding to $t$-channel gauge boson
exchange between two scalars.  Production cross sections that contain
the CP-odd Higgs or the charged Higgs instead decrease like $1/s$,
because no scalar-Goldstone-gauge boson vertices exist for these
particles.

\smallskip

In the cases involving first and second generation sfermions,
$t$-channel sfermion exchange with an initial-state $W$ contributes
only to left-handed sfermions, so the $\tilde f_L\tilde f_L^*$ cross
sections dominate over $\tilde f_R\tilde f_R^*$.  In the neutralino
sector, $\neua$ is dominantly bino and does not couple to neutral
Higgs bosons, so $\neua$ production in $W^+W^-$ fusion is suppressed.
The other neutralinos and charginos, being the SUSY partners of
massive vector bosons and Higgses, are produced with cross sections up
to $100\;\pb$.  The largest neutralino rates occur for mixed gaugino
and higgsino production, because the Yukawa couplings are given by the
gaugino--higgsino mixing entry in the neutralino mass matrix.  In the
Higgs sector, the decoupling limit ensures that only $W^+W^-\to Zh$,
$WZ\to Wh$ (almost $100\;\pb$), and $W^+W^-\to hh$ ($6\;\pb$) are
important, while the production of heavy Higgses is suppressed.  For
$W^+W^-\to Ah$ and $W^-Z\to H^-h$ the decoupling suppression applies
twice.

\medskip

In reality, $WW\to XX$ and $WZ\to XY$ scattering occurs only as a
subprocess of $2\to 6$ multi-particle production.  The initial vector
bosons are emitted as virtual states from a pair of incoming fermions.
The measurable cross sections are phase-space suppressed by a few
orders of magnitude.  A rough estimate can be made by folding the
energy-dependent $WW/WZ$ cross sections with weak-boson structure
functions.  Reliable calculations require the inclusion of all Feynman
diagrams, as can be done with the programs presented in this paper ---
the production rates rarely exceed ${\cal O}(\ab)$ at the
LHC~\cite{WBF}.


\subsubsection{Other processes}

For the remaining lists of processes with vector-boson or fermion
initial states, similar considerations apply.  In particular, the
photon has no longitudinal component, so $\gamma$-induced electroweak
processes (Tables~\ref{xsec:W-A},~\ref{xsec:ZA} and~\ref{xsec:AA}) are
not related to Goldstone-pair scattering.  We include unrealistic
fermionic initial states such as $\tau^+\tau^-$,
$\tau^-\bar{\nu_\tau}$ and $b\bar{t}$ in our reference list,
Tables~\ref{xsec:tau+tau-},~\ref{xsec:tau-nu} and~\ref{xsec:bt},
because they involve Feynman rules that do not occur in other
production processes, but are relevant for decays.

Finally, our set of processes contains several lists with the colored
fermionic initial states $u\bar{u}$, $d\bar{d}$ and $b\bar{b}$
(Tables~\ref{xsec:uu}--\ref{xsec:bb}, plus Table~\ref{xsec:ff-iden}
for same-flavor fermions); $gg$-fusion (Table~\ref{xsec:GG});
$qg$-fusion (Table~\ref{xsec:qG}); and mixed QCD-electroweak processes
$gA$, $gZ$ and $gW$ (Tables~\ref{xsec:GA},~\ref{xsec:GZ}
and~\ref{xsec:GW-}).  These (as full hadronic processes) are
accessible at hadron colliders, and comparing their cross sections
completes our check of Feynman rules of the SUSY-QCD sector and its
interplay with the electroweak interactions.  Note that for a
transparent comparison we do not fold the quark- and gluon-induced
processes with structure functions.

\medskip

The only Feynman rules not checked by any process in this list are the
four-scalar couplings.  It is expected, and has explicitly been
verified for the four-Higgs coupling in
particular~\cite{quartic_higgs}, that these contact interactions are
not accessible at any collider in the foreseeable future.  We
therefore neglect them.


\subsection{Flavor Mixing}
\label{sec:ckm}

For most of this paper we have neglected the quark masses and the
mixings of the first two squark and slepton/sneutrino generations.
Here we give a brief account of the consequences of using a
non-diagonal CKM matrix.  Full CKM mixing is available as an option
for the {\sc whizard} and {\sc sherpa} event generators.  For {\sc
madgraph}, it is straightforward to modify the model definition file
accordingly.

\smallskip

The CKM mixing matrix essentially drops out from most processes when
we sum over all quark intermediate and final states.  This is due to
CKM unitarity, violated only by terms proportional to the quark mass
squared over $\sqrt{s}$ in high energy scattering processes.  For the
first two generations, such corrections are negligible at the energies
we are considering.

At hadron colliders, summation over initial-state flavors does not
lead to cancellation because the parton densities are
flavor-dependent.  In the SM, CKM structure matters only for
charged-current processes where a $q\bar{q}'$ pair annihilates into a
$W$ boson.  For instance, the cross section for $u\bar{d}\to W^{+*}\to
X$ is multiplied by $|V_{ud}|^2$, and the cross section for
$u\bar{s}\to W^{+*}\to X$ is proportional to $|V_{us}|^2$.

In the partonic final state, CKM unitarity ensures that a cross
section does not depend on flavor mixing.  However, jet hadronization
depends on the jet quark flavor.  Neglecting CKM mixing can result in
a wrong jet-flavor decomposition.  In practice, this is not relevant
since jet-flavor tagging (except for $b$ quarks, and possibly for $c$
quarks) is impossible.  In cases where it is relevant, \eg charm
tagging in Higgs decay backgrounds at an ILC, the problem may be
remedied either by reverting to the full CKM treatment, or by rotating
the outgoing quark flavors before hadronization on an event-by-event
basis.

\medskip

To estimate the impact of CKM mixing on SUSY processes, we consider
the electroweak production of two light-flavor squarks at the LHC:
$q\bar{q}\to\tilde{q}'\tilde{q}'{}^*$.  Adopting the input of
Appendix~\ref{sec:input} and standard values for the CKM mixing
parameters reduces the cross section by about $4\%$,
Tab.~\ref{tab:ckm}.  This is negligible for LHC phenomenology, but
ensures a correct implementation of CKM mixing in the codes.

\begin{table}
\begin{displaymath}
  \begin{array}{|l|r|}
    \hline
    \multicolumn{2}{|c|}{\text{CKM diagonal}}
    \\
    \hline
    u\bar u \to \sdl\sdl^* & 166.621(8)
    \\
    u\bar u \to \ssl\ssl^* & 175.686(9)
    \\
    \hline
    d\bar d \to \sul\sul^* & 174.678(9)
    \\
    d\bar d \to \scl\scl^* & 178.113(9)
    \\
    \hline
  \end{array}
\hspace{15mm}
  \begin{array}{|l|r|}
    \hline
    \multicolumn{2}{|c|}{\text{with CKM}}
    \\
    \hline
    u\bar u \to \sdl\sdl^* & 160.547(8)
    \\
    u\bar u \to \ssl\ssl^* & 168.733(8)
    \\
    \hline
    d\bar d \to \sul\sul^* & 167.875(8)
    \\
    d\bar d \to \scl\scl^* & 170.984(9)
    \\
    \hline
  \end{array}
\end{displaymath}
\caption{Squark production cross sections computed using 
  {\sc sherpa}/{\sc whizard} with and without non-trivial CKM mixing.
\label{tab:ckm}}
\end{table}

Finally, there can be nontrivial flavor effects in the soft
SUSY-breaking parameters.  That is, if squark mixing differs from
quark mixing, in the case of flavor-dependent SUSY
breaking~\cite{flavor_review}.  Non-minimal flavor violation predicts
large signals for physics beyond the Standard Model, in particular
flavor-changing neutral currents, in low-energy precision observables
like kaon mixing.  Their absence is a strong indication of flavor
universality in a SUSY breaking mechanism.  However, if desired,
nontrivial SUSY flavor effects can be included by the codes with minor
modifications.


\section{Sbottom Production at the LHC}
\label{sec:lhc}

A SUSY process of primary interest at the LHC is bottom squark
production.  For this specific discussion, we adopt a SUSY parameter
point with rather light sbottoms and a rich low-energy phenomenology.
The complete parameter set is listed in Appendix~\ref{sec:bpoint}.
The sbottom masses are
\begin{equation}
  m_{\sbl} = 295.36\,\GeV,
\qquad
  m_{\sbr} = 399.92\,\GeV.
\end{equation}
In the following we will focus on the decay $\sbl\to b\neua$ with a
branching ratio of $43.2\%$.  The lightest Higgs boson is near the LEP
limit, but decays invisibly to neutralinos with a branching ratio of
$44.9\%$.  The heavy Higgses are at $300\,\GeV$.  The lightest
neutralino mass is $m_{\neua}=46.84\,\GeV$, while the other
neutralinos and charginos are between $106$ and $240\,\GeV$.  Sleptons
are around $200\,\GeV$.  The squark mass scale is $430\,\GeV$ (except
for $m_{\str}$), and the gluino mass is $800\,\GeV$.

A spectacular signal at this SUSY parameter point would of course be
the light Higgs. Apart from SUSY decays, our light MSSM Higgs sits in
the decoupling region, which means it is easily covered by the MSSM
No-Lose theorem at the LHC~\cite{no_lose}: for large pseudoscalar
Higgs masses a light Higgs will be seen by the Standard Model searches
in the WBF $\tau\tau$ channel.  Unfortunately, in most scenarios it
would be challenging to distinguish a SUSY Higgs boson from its SM
counterpart, after properly including systematic errors.  Here, our
SUSY parameter point predicts a large light Higgs boson invisible
branching fraction, which would also be visible in the WBF
channel~\cite{invisible_higgs}.  There would be little doubt that this
light Higgs is not part of the SM Higgs sector.

\medskip

We have checked that our SUSY parameter point satisfies the low-energy
constraints for $\Delta\rho$~\cite{delta_rho_ex,delta_rho_th},
$g_\mu-2$~\cite{gminus2_ex,gminus2_th}, $b\to
s\gamma$~\cite{bsgamma_ex,bsgamma_th} and
$B_s\to\mu^+\mu^-$~\cite{bsmumu_ex,bsmumu_th}, as well as the
exclusion limits for Higgs and SUSY particles.  The relic neutralino
density~\cite{wmap_th} is below the observed dark-matter
density~\cite{wmap_ex} and therefore allowed.

\medskip

While this point might look slightly exceptional, in particular
because of the large invisible light Higgs branching ratio, the only
parameters which matter for sbottom searches at the LHC are the fairly
small sbottom masses.  The current direct experimental limits come
from the Tevatron search for jets plus missing energy, where at least
for CDF the jets include bottom quark tags~\cite{tevatron}.  However,
for sbottom production the Tevatron limit has to be regarded as a
limit on cross section times branching ratio.  The mass limits derived
in the light-flavor squark and gluino mass plane assume squark pair
production including diagrams with a $t$-channel gluino, which is
strongly reduced for final-state sbottoms.  Moreover, strong mass
limits arise from associated squark--gluino production, which is also
largely absent in the case of sbottoms~\cite{prospino}.

Searching for squark and gluino signatures at the LHC as a sign of
physics beyond the Standard Model (such as SUSY) has one distinct
advantage: once we ask for a large amount of missing energy, the
typical SM background will involve a $W$ or $Z$ boson.  Because
squarks and gluinos are strongly interacting, the signal-to-background
ratio $S/B$ is automatically enhanced by a factor $\alpha_s/\alpha$.
This means that for typical squark and gluino masses below ${\cal
O}(\TeV)$ we expect to see signs of new physics before we see a
light-Higgs signal.  Most SUSY mass spectrum information is carried by
the squark and gluino cascade decay
kinematics~\cite{susy_spins,cascade}, and we are confident that,
though non-negligible, QCD effects will not alter these results
dramatically~\cite{skands}.  The most dangerous backgrounds to cascade
decay analyses are not SM $Z$+jets events, but SUSY backgrounds, for
example simple combinatorics with two decay chains in the same event.
The (less likely) case that SUSY particles are produced at the LHC,
but do not decay within the detector, is an impressive show of the
power of the LHC detectors --- finding and studying these particles
does not pose a serious problem at either ATLAS or
CMS~\cite{lhc_stable}.


\subsection{Off-Shell Effects in Sbottom Decays}

\begin{figure}[t]
\begin{center}
\includegraphics[scale=1.00]{bbnn.3} \hspace*{15mm}
\includegraphics[scale=1.00]{bbnn.4}
\end{center}
\vspace*{4mm}
\caption{The $p_{T,b}^{\rm max}$ (left) and $\met$ (right)
  distributions for the signal process $gg\to b\bar{b}\neua\neua$ and
  the main SM background $pp\to b\bar{b}\nu\bar\nu$, at the LHC.  The
  missing transverse momentum $\met$ is defined as the transverse
  momentum of the $\neua\neua$ or $\nu\bar{\nu}$ pair and does not
  include $b$ decay products.  Both processes are evaluated including
  all off-shell diagrams.}
\label{fig:lhc_back}
\end{figure}
\begin{figure}[t]
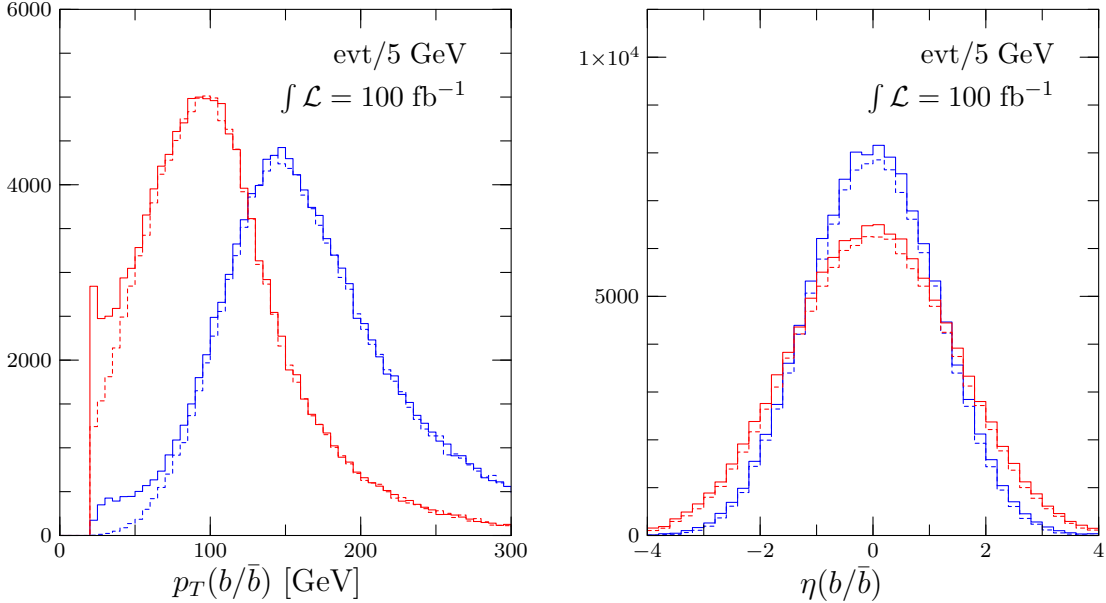

\begin{center}
\includegraphics[scale=1.00]{bbnn.1} \hspace*{15mm}
\includegraphics[scale=1.00]{bbnn.2}
\end{center}
\vspace*{4mm}
\caption{The $p_{T,b}$ (left) and $\eta_b$ (right) distributions for 
  $gg\to b\bar{b}\neua\neua$ at the LHC.  The blue (red) curves
  correspond to the harder (softer) of the two $b$ jets.  The dashed
  lines show the Breit-Wigner approximation for sbottoms; solid lines
  include all off-shell effects.}
\label{fig:lhc_bw}
\end{figure}

From a theoretical point of view, the production process
$pp\to\sbl\sbl^*$ with subsequent dual decays $\sbl\to b\neua$ can be
described using two approximations.  Because the sbottoms are scalars,
their production and decay matrix elements can be separated by an
approximate Breit-Wigner propagator.  Furthermore, the sbottom width
$\Gamma_{\tilde{b}_1}=0.53$~GeV is sufficiently small to safely assume
that even extending the Breit-Wigner approximation to a narrow-width
description should result in percent-level effects, unless cuts force
the sbottoms to be off-shell.

\medskip

For this entire LHC section we require basic cuts for the bottom
quark, whether it arises from sbottom decays or from QCD jet
radiation: $p_{T,b}>20$~GeV and $|\eta_b|<4$.  We require any two
bottom jets to be separated by $\Delta R_{bb}>0.4$.  There are no
additional cuts, for example on missing transverse energy, because we
do not attempt a signal vs. background analysis.  Instead, we focus on
the approximations which enter the signal process calculation.

To stress the importance of properly understanding the signal process'
distributions, we show those for $p_{T,b}^{\rm max}$ and $\met$ for
the signal process $gg\to b\bar{b}\neua\neua$ and for the main SM
background $pp\to b\bar{b}\nu\bar\nu$ in Fig.~\ref{fig:lhc_back}.  As
expected, all final-state particles are considerably harder for the
signal process.  This is due to heavy intermediate sbottoms in the
final state.  Historically, these kinds of distributions for QCD
backgrounds have played an important role illustrating progress in the
proper description of jet radiation, a discussion we turn to in the
next section.  The $\met$ distribution is only a parton-level
approximation, \ie the transverse momentum of the $\neua\neua$ or
$\nu\bar\nu$ pair and does not include $b$ decays.  However, we expect
the $b$-decay contributions to be comparably small and largely
balanced between the two sbottom decays.

The effects of the Breit-Wigner approximation compared to the complete
set of off-shell diagrams are shown in Fig.~\ref{fig:lhc_bw}.  After
basic cuts the cross section for the process $gg\to\sbl\sbl^*\to
b\bar{b}\neua\neua$ is 1120~fb.  Because of the roughly 250~GeV mass
difference between the decaying sbottom and the final-state
neutralino, even the softer $b$ jet $p_T$ distribution peaks at
100~GeV.  As expected from phase space limitations, the harder of the
$b$ jets is considerably more central, but for both of the final-state
bottom jets an additional tagging-inspired cut $|\eta_b|<2.5$ would
capture most events.  Including all off-shell contributions, \ie
studying the complete process $gg \to b\bar{b}\neua\neua$, leads to a
small cross section increase, to 1177~fb after basic cuts.  The
additional events are concentrated at softer jet transverse momenta
($p_{T,b}\lesssim 60$~GeV) and alter the shape of the distributions
sizeably.  The diagrams which can contribute to off-shell effects are,
for example, bottom quark pair production in association with a
slightly off-shell $Z$, where the $Z$ decays to two neutralinos.  The
remaining QCD process $gg\to b\bar{b}$ produces much softer $b$ jets,
because of the lack of heavy resonances.  Luckily, this considerable
distribution shape change is mostly in a phase space region plagued by
large background, as shown in Fig.~\ref{fig:lhc_back}, therefore will
be removed in an analysis.  On the other hand, there is no guarantee
that off-shell effects will always lie in this kind of phase space
region, and in Fig.~\ref{fig:lhc_bw} we can see that the Breit-Wigner
approximation is by no means perfect.


\subsection{Bottom-Jet Radiation}

\begin{figure}[t]
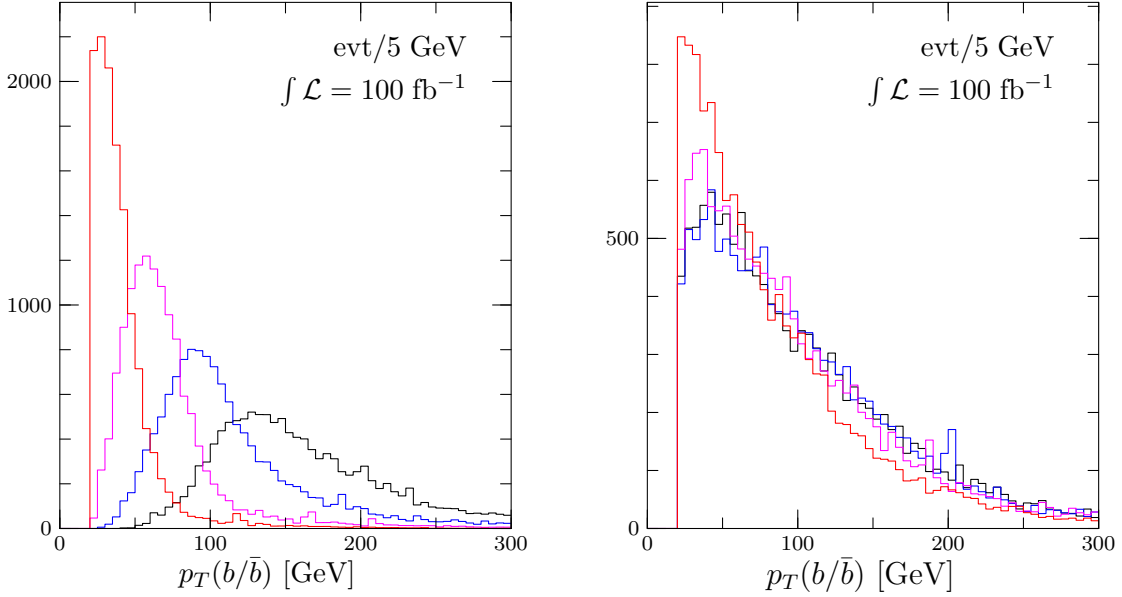

\begin{center}
\includegraphics[scale=1.00]{bbnn.5} \hspace*{15mm}
\includegraphics[scale=1.00]{bbnn.6}
\end{center}
\vspace*{4mm}
\caption{The $p_{T,b}$ distributions for the LHC process    
  $gg\to b\bar{b}b\bar{b}\neua\neua$.  The left panel orders the jets
  according to their $p_{T,b}$, while in the right panel they are
  ordered by $|\eta_b|$.  These peaks from left to right corresond to
  more central jets.}
\label{fig:lhc_pt}
\end{figure}
\begin{figure}[t]
\begin{center}
\includegraphics[scale=1.00]{bbnn.7} \hspace*{15mm}
\includegraphics[scale=1.00]{bbnn.8}
\end{center}
\vspace*{4mm}
\caption{The $p_{T,b}^{\rm max}$ (left) and $\met$ (right) 
  distributions for $gg\to b\bar{b}b\bar{b}\neua\neua$ (red) and 
  $gg\to b\bar{b}\neua\neua$ (blue) at the LHC.}
\label{fig:lhc_2b4b}
\end{figure}

Just as with light-flavor squarks in $q\bar{q}$ scattering, LHC could
produce sbottom pairs from a $b\bar{b}$ initial state.  Bottom
densities~\cite{acot} and SUSY signatures at the LHC are presently
undergoing careful study~\cite{berdine}.  However, for heavy Higgs
production it was shown that bottom densities are the proper
description for processes involving initial-state bottom quarks.  The
comparison between gluon-induced~\cite{higgs_gluon} and
bottom-induced~\cite{higgs_bottom} processes backs the bottom-parton
approach, as long as the bottom partons are defined
consistently~\cite{eduard_bottom}.  The bottom-parton picture for
Higgs production becomes more convincing the heavier the final state
particles are~\cite{bottom_heavy}, \ie precisely the kinematic
configuration we are interested in for SUSY particles~\cite{berdine}.

\medskip

Sbottom pair production is the ideal process for a first attempt to
study the effects of bottom jet radiation on SUSY-QCD signatures.  In
the fixed-flavor scheme (only light-flavor partons) the leading-order
production process for sbottom pairs is $2\to 2$ gluon fusion.  If we
follow fixed-order perturbation theory, the radiation of a jet is part
of the NLO corrections~\cite{prospino}.  This jet is likely to be an
initial-state gluon, radiated off the $gg$ or $q\bar{q}$ initial
states.  Crossing the final- and initial-state partons, $qg$
scattering would contribute to sbottom pair production at NLO, adding
a light-flavor quark jet to the final state.  The perturbative series
for the total rate is stable, and as long as the additional jet is
sufficiently hard ($p_{T,j}\gtrsim 50$~GeV), the ratio of the
inclusive cross sections is small:
$\sigma_{\tilde{b}\tilde{b}j}/\sigma_{\tilde{b}\tilde{b}}\sim
1/3$~\cite{skands}.

With the radiation of two jets (at NNLO in the fixed-flavor scheme),
the situation becomes more complicated.  We know that QCD jet
radiation at the LHC is not necessarily softer than jets from SUSY
cascade decays~\cite{skands}.  This jet radiation can manifest itself
as a combinatorial background in a cascade analysis.  Here we study
the energy spectrum of bottom jets from the decay $\sbl\to b\neua$, so
additional bottom jets from the initial state lead to combinatorial
background.  Once we radiate two jets from the dominant $gg$ initial
state, bottom jets appear as initial-state radiation (ISR).  In the
total rate this process can be included just by using the
variable-flavor scheme in the leading-order cross section, as
discussed above.

\smallskip

As expected, the rate for the production process $gg\to
b\bar{b}b\bar{b}\neua\neua$ of 130.7~fb is considerably suppressed
compared to the 1177~fb for inclusive (off-shell) sbottom pair
production.  Again, we require $p_{T,b}>20$~GeV.  The $b$-jet
multiplicity is expected to decrease once we require harder $b$-jets
in a proper analysis.  The reduction factor for two additional bottom
jets is $\sim 1/3\times 1/3$, as quoted above from Ref.~\cite{skands}
for general jet radiation.  However, we include considerably softer
$b$ jets as compared to the 50~GeV light-flavor jets which lead to a
similar reduction factor.  The reason is that high-mass final states
at the LHC are most efficiently produced in quark-gluon scattering,
and in our analysis we are limited to gluons for both incoming
partons.

From our more conceptual point of view, the crucial question is how to
identify the decay $b$ jets, which carry information on the SUSY mass
spectrum~\cite{cascade}.  Because the ISR $b$ jets arise from gluon
splitting, they are predominantly soft and forward in the detector.
To identify the decay $b$ quarks we can try to exclude the most
forward and softest of the four $b$ jets in the event, to reduce the
combinatorial background.  In Fig.~\ref{fig:lhc_pt} we show the
ordered $p_{T,b}$ spectra of the four final-state sbottoms.  Because
of kinematics we would expect that it should not matter if we order
the sbottoms according to $p_{T,b}$ or $|\eta_b|$, at least for
grouping into initial-state and decay jet pairs.  However, we see that
this kinematical argument is not well suited to remove combinatorial
backgrounds.  Only the most forward $b$ jet is indeed slightly softer
than the other three, but the remaining three $p_{T,b}$ distributions
ordered according to $|\eta_b|$ are indistinguishable.

\smallskip

After discussing the combinatorial effects of additional $b$ jets in
the final state, the important question is whether additional $b$-jet
radiation alters the kinematics of sbottom production and decay.  In
Fig.~\ref{fig:lhc_2b4b} we show the $p_{T,b}^{\rm max}$ and the $\met$
distributions for $b\bar{b}\neua\neua$ and
$b\bar{b}b\bar{b}\neua\neua$ production at the LHC; those most likely
to be useful in suppressing SM backgrounds.  The soft ends of the
$p_{T,b}$ distributions do not scale because in the $4b$ case the
hardest $b$ jet becomes less likely to be a decay $b$-jet.  Instead, a
soft decay $b$ quark will be replaced with a harder initial-state $b$
jet in our distribution.  The $4b$ distribution peaks at lower
$p_{T,b}$ because the minimum cut on $p_{T,b}$ of the
initial-state $b$ jets eats into the steep gluon densities.  At very
large values of $p_{T,b}$ this effect becomes relatively less
important, and the two distributions scale with each other.

The $\met$ distributions, however, are sensitive to $\sum p_{T,b}$.
If both $b$-jets come from heavy particle decays, the decay can alter
their back-to-back kinematics.  In contrast, additional light particle
production balances out the event, leading to generally smaller $\met$
values.  We might be lucky in the final analysis, because a proper
analysis after background rejection cuts will be biased toward small
$\met$, thus will be less sensitive to $b$-jet radiation and
combinatorial backgrounds.


\section{Sbottom Production at an ILC}
\label{sec:ilc}

At an ILC we would be able to obtain more accurate mass and cross
section measurements, provided the collider energy is sufficient to
produce sbottom pairs.  This is due to the much cleaner lepton
collider environment, relative to a hadron collider -- even though the
lower rate can statistically limit measurements.  For this study we
again choose the parameter point described in
Appendix~\ref{sec:bpoint}.  There, the sbottom mass is low, but the
appearance of various Higgs and neutralino backgrounds complicates the
analysis.

With sbottom production we encounter a process where multiple channels
and their interferences contribute to the total signal rate; this is
more typical than not.  We are forced to understand off-shell effects
to perform a sensible precision analysis.  Assuming $800\;\GeV$
collider energy, the production channels $\sbl\sbl^*$ and $\sbl\sbr^*$
are open.  From the squark-mixing matrix it can be seen that the
lighter of the two sbottoms, $\sbl$, predominantly is right-handed.
Its main decay mode is to $b\neua$.  Therefore, as with sbottom
production at the LHC, the principal final state to be studied is
$b\bar{b}$ plus missing energy.

At the LHC, sbottom pair production dominates this final state because
it is the only strongly-interacting production channel.  In contrast,
sbottom pair production at an ILC would proceed via electroweak
interactions.  Hence, all electroweak SUSY and SM processes that
contribute to the same final state need to be considered.  In
particular, the following $2\to 2$ production processes contribute to
$e^+e^-\to b\bar{b}\neua\neua$:
\begin{align}
e^+e^- & \to Zh,\ ZH,\ Ah,\ AH,\ \neua\neub,\ \neua\neuc,\ \neua\neud,\
             \sbl\sbl^*,\ \sbl\sbr^* \; .
\end{align}
All cross sections, in different approximations as well as in a
complete calculation including all interferences, are displayed in
Table~\ref{tab:channels}.  Once we fold in the branching ratios, fewer
processes contribute significantly, namely:
\begin{align}
e^+e^- & \to Zh,\ AH,\ \neua\neub,\ \neua\neuc,\
             \sbl\sbl^*,\ \sbl\sbr^* \; .
\end{align}
The SM process $e^+e^-\to b\bar{b}\nu_i\bar\nu_i \; (i=e,\mu,\tau)$ is
dominated by $WW$ fusion to $Z/h$ (followed by $Z/h\to b\bar{b}$) and
by $Zh/ZZ$ pair production. It represents a significant irreducible
background, as a neutrino cannot be distinguished from the lightest
neutralino in high-energy collisions.  Thus, we refer to this final
state with neutrinos as the SM background.

\begin{table}
\begin{displaymath}
  \begin{array}{|l|rrr|}
    \hline
    \text{Channel} & \sigma_{2\to 2}~[\fb] & 
    \sigma\times\mathrm{BR}~[\fb] & \sigma_{\rm BW}~[\fb]
    \\
    \hline
    Zh           & 20.574 & 1.342 & 1.335
    \\
    ZH           &  0.003 & 0.000 & 0.000
    \\
    hA           &  0.002 & 0.001 & 0.000
    \\
    HA           &  5.653 & 0.320 & 0.314
    \\
    \neua\neub   & 69.109 &13.078 &13.954
    \\
    \neua\neuc   & 24.268 & 3.675 & 4.828
    \\
    \neua\neud   & 19.337 & 0.061 & 0.938
    \\
    \sbl\sbl     &  4.209 & 0.759 & 0.757
    \\
    \sbl\sbr     &  0.057 & 0.002 & 0.002
    \\
    \hline
    \text{Sum}   &        &19.238 &22.129
    \\
    \hline\hline
    \text{Exact} &        &       &19.624
    \\
    \text{w/ISR}         &&       &22.552
    \\
    \hline
  \end{array}
\hspace*{5mm}
  \begin{array}{|l|rrr|}
    \hline
    \text{Channel} & \sigma_{2\to 2/3}~[\fb] & 
    \sigma\times\mathrm{BR}~[\fb] & \sigma_{\rm BW}~[\fb]
    \\
    \hline
    ZZ           & 202.2 &  12.6 &  13.1
    \\
    Zh           &  20.6 &   1.9 &   1.9
    \\
    ZH           &   0.0 &   0.0 &   0.0
    \\
    \hline\hline
    Z\bar\nu\nu  & 626.1 & 109.9 & 111.4
    \\
    h\bar\nu\nu  & 170.5 &  76.5 &  76.4
    \\
    H\bar\nu\nu  &   0.0 &   0.0 &   0.0
    \\
    \hline
    \text{Sum}   &       & 186.5 & 187.7
    \\
    \hline\hline
    \text{Exact} &       &       & 190.1
    \\
    \text{w/ISR} &       &       & 174.2
    \\
    \hline
  \end{array}
\end{displaymath}
\caption{SUSY cross sections contributing to 
  $e^+e^-\to b\bar{b}\neua\neua$ (left) and the SM background
  $e^+e^-\to b\bar{b}\nu\bar\nu$ (right).  The columns assume:
  on-shell production; same, including the branching ratio into
  $b\bar{b}\neua\neua$ and $b\bar{b}\nu\bar\nu$; and with a
  Breit-Wigner propagator.  The incoherent sum is shown at the bottom.
  In the SM case, only the $2\to 3$ processes are summed, to avoid
  double-counting.  The exact tree-level result includes all Feynman
  diagrams and interferences.  The last line shows the effect of
  initial-state radiation (ISR) and beamstrahlung.
\label{tab:channels}}
\end{table}
%


\subsection{Numerical Approximations}

It is instructive to compare various levels of approximation found in
the literature before moving to a complete treatment of the process.
The simplest approximation for resonant production and decay is to
multiply the production cross section by the appropriate branching
fraction.  This narrow width approximation (NWA) is expected to hold
as long as $\Gamma/m \ll 1$.  In traditional Monte Carlos, angular
correlations are lost for scalar resonances unless spin correlations
along the lines of Ref.~\cite{Richardson:2001df} are included.

We can improve upon this by constraining the intermediate state to
resonances (in our case the two sbottoms) and inserting Breit-Wigner
propagators.  Such an approach takes into account off-shell
corrections that originate from the nontrivial resonance kinematics.
However, the Breit-Wigner amplitude is not gauge-invariant
off-resonance, thus the precise result depends on the choice of gauge
(unitarity gauge in our calculations).  Both, this approximation and
the NWA neglect interferences with off-resonant diagrams.

To obtain the full tree-level result, all Feynman graphs and their
interferences must be taken into account, and an unambiguous breakdown
into resonance channels is no longer possible.  Perturbation theory
breaks down at the poles of intermediate on-shell states.  The
emerging divergences have to be regularized, for example via finite
particle widths which unitarize the amplitude.  Not surprisingly,
na\"ively including particle widths violates gauge invariance, but
schemes exist which properly address this
problem~\cite{LEP2-Gauge-Gospel}.  All our codes use the fixed-width
scheme, which includes the finite width even in the spacelike region
and avoids problems of gauge invariance in the processes we consider
here.

Finally, in many cases the effects of initial-state radiation (ISR)
and beamstrahlung are numerically of the same order of magnitude as
the full resonance and interference corrections, or even larger, and
therefore need to be addressed.


\subsection{Particle Widths}

As discussed before, we must include finite widths for all
intermediate particles that can become on-shell.  For the processes
discussed here this includes the neutral Higgs and $Z$ bosons, the
neutralinos, and the sbottoms.  It is tempting to merely treat the
widths as externally fixed numerical parameters.  This, however, can
lead to a mismatch: consider a tree-level process with an intermediate
resonance with mass $M$ and total width $\Gamma$.  The tree-level
cross section contains a factor
\begin{equation}\nonumber
\frac{1}{(p^2-M^2)^2+M^2\Gamma^2} \; .
\end{equation}
In the vicinity of the pole a factor $1/\Gamma$ is picked up.  If
$\Gamma\ll M$, this contribution to the cross section can be
approximated by the on-shell production cross section multiplied by
the branching ratio for resonance decay into the desired final state
$X$, \ie $\mathrm{BR}_X=\Gamma_X/\Gamma$ (cf.\ Sec.~\ref{sec:multi}).
While the total width $\Gamma$ is an external numerical parameter, the
partial width $\Gamma_X$ is implicitly computed by the integration
program at tree-level during cross section evaluation.  This can lead
to a noticeable mismatch, especially if the external full width is
calculated with higher-order corrections. Formally, the use of
loop-improved widths induces an order mismatch in any leading-order
calculation, which, in principle, is allowed.  However, in reality,
dominant corrections might reside in both the decay (width)
calculation and the production process, canceling each other in the
full result.  The NLO corrections to the full process that would
remedy the problem are generally unavailable, at least in a form
suitable for event generation~\cite{krr}.  To illustrate this
reasoning, consider a case where the resonance has only one decay
channel.  Then, in the narrow-width limit, the factorized result is
reproduced only if the tree-level width is taken as an input computed
from exactly the same parameters as the complete process.

While this looks like a trivial requirement, it should be stressed
that most MSSM decay codes return particle widths that include higher
orders, either explicitly or implicitly through the introduction of
running couplings and mass parameters.  Similarly, for the $Z$ boson
width one is tempted to insert the measured value, which in the best
of all worlds corresponds to the all-orders perturbative result.  To
avoid the problems mentioned above, in this paper we calculate all
relevant particle widths in the same tree-level framework used for the
full process.  For completeness, we list them in Tab.~\ref{tab:decay}
of Appendix~\ref{sec:bpoint}, corresponding to the SLHA input file
used for the collider calculation.  Our leading-order widths agree
with those of {\sc sdecay}~\cite{sdecay}.


\subsection{Testing the Narrow Width Approximation}

\begin{figure}
\begin{center}
\includegraphics[scale=0.8]{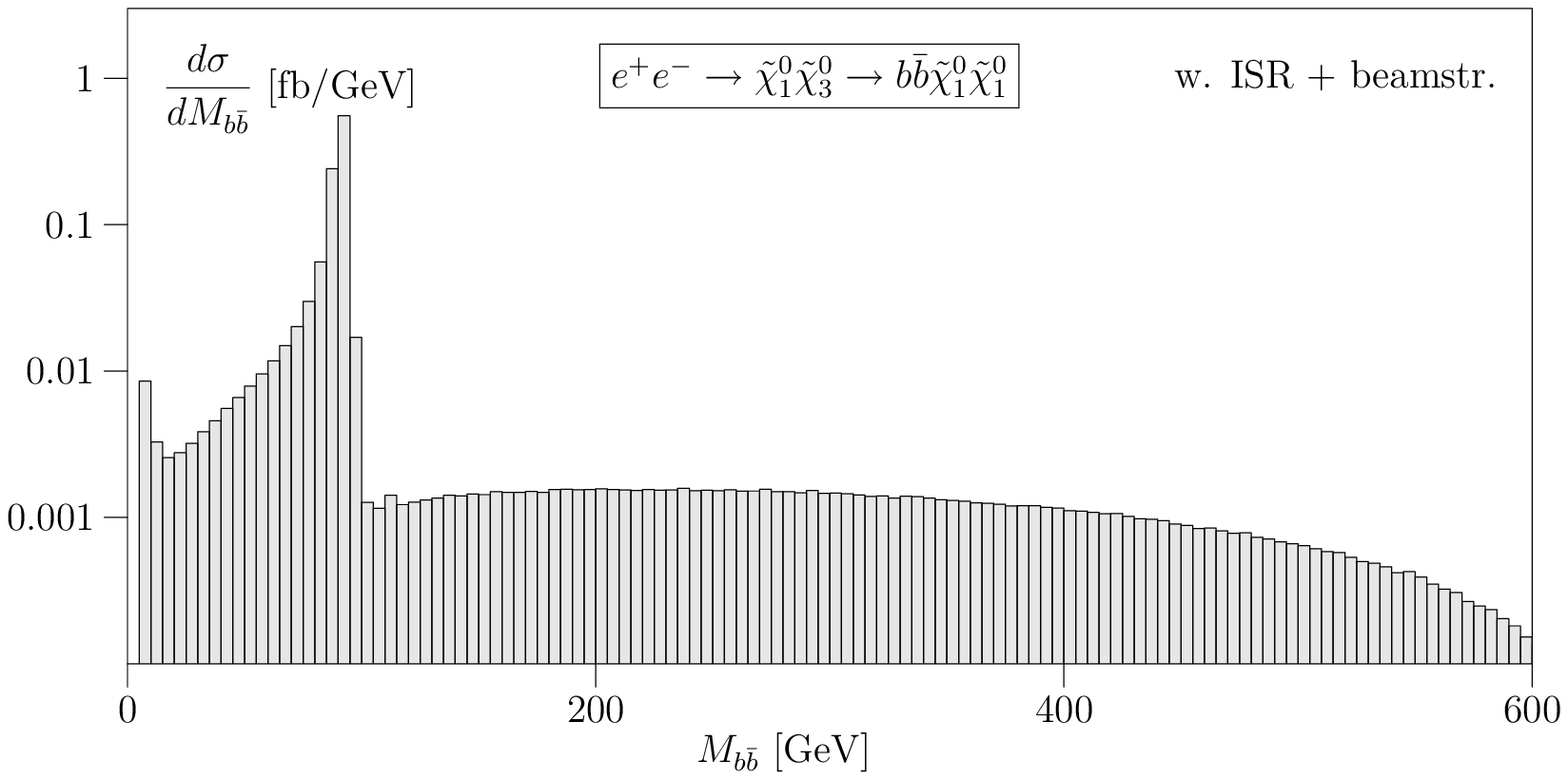}
\end{center}
\vspace*{-8mm}
\caption{The $b\bar{b}$ invariant mass distribution for the $\neua\neuc$
  contribution to $e^+e^-\to b\bar{b}\neua\neua$.}
\label{fig:neu1neu3}
\end{figure}

An estimate of the effects of the NWA and of Breit-Wigner propagators
is shown in Tab.~\ref{tab:channels}.  In replacing on-shell
intermediate states by Breit-Wigner functions in the SUSY processes
(left panel) the total cross section increases by $15\%$.  Breaking
the cross section down into individual contributions, it becomes
apparent that this increase is mainly due to the heavy neutralino
channels.  In contrast, the $Z$, Higgs and sbottom channels are fairly
well-described by the on-shell approximation of
Eq.~(\ref{eq:on-shell}).  Including the complete set of all tree-level
Feynman diagrams with all interferences results in a decrease of
$11\%$.  Obviously, continuum and interference effects are
non-negligible and must be properly taken into account.

Similar considerations apply to the SM background, $e^+e^-\to
b\bar{b}\nu\bar\nu$, shown in the right panel of
Tab.~\ref{tab:channels}.  At a collider energy of $800\;\GeV$, the SM
process is dominated by weak boson fusion, while pair production
($ZZ/ZH$) borders on negligible.  For the total cross section, the NWA
works well: inserting Breit-Wigner propagators for the intermediate
$Z,h$ states increases the rate by a mere $0.6\%$, and including all
diagrams with interferences leads to a further increase of only
$1.3\%$.

\smallskip

Finally, we compute the effect of ISR and beamstrahlung: the SUSY
cross section increases by $15\%$ --- a general effect seen for
processes dominated by particle pair production well above threshold.
(In that range the cross sections are proportional to $1/\hat{s}$ and
therefore profit from the reduction in effective energy due to photon
radiation.)  In contrast, for the SM background, adding ISR and
beamstrahlung amounts to a reduction by $8\%$.  This is expected for a
$t$-channel-dominated process with asymptotically flat energy
dependence.

\medskip

Apart from total cross sections, it is crucial to understand off-shell
effects in distributions.  They are significant in the neutralino
channels $e^+e^-\to\neua\tilde{\chi}_i^0\;(i=2,3,4)$, the dominant
SUSY backgrounds to our sbottom signal.  For this mass spectrum, the
$\neub$ has a three-body decay to $q\bar{q}\neua$; here the focus is
on $q=b$.  The higgsino-like $\neuc$ has a two-body decay $\neuc\to
Z\neua$ with a branching fraction close to $100\%$~\cite{sdecay}.

In the complete calculation, neither the decaying $\neuc$ nor the
intermediate $Z$ is forced on-shell.  Continuum effects play a role.
This explains the differences in the decay spectrum between the full
calculation and the approximation using Breit-Wigner propagators, as
seen in Fig.~\ref{fig:neu1neu3}.  There, we include neutralino pair
production, $e^+e^-\to\neua\neuc$.  In Fig.~\ref{fig:neu1neu3} we show
the $b\bar{b}$ invariant mass spectrum for the process
$e^+e^-\to\neua\neuc\to b\bar{b}\neua\neua$.  Assuming a two-body
$\neuc$ decay, one would expect a sharp Breit-Wigner $Z$ resonance at
$91.18\;\GeV$.  Instead, the resonance is not Breit-Wigner-like and is
surrounded by a nearly flat continuous distribution at both high and
low masses.  Clearly, this would not be accounted for by a factorized
production--decay approximation.  In fact, it stems from a highly
off-shell three-body decay $\neuc\to b\bar{b}\neua$ via an
intermediate sbottom.  As a background to sbottom pair production,
this process gives the dominant contribution, because we can easily
cut against on-shell neutralino production.  The significant low-mass
tail explains the $30\%$ enhancement for this channel seen in
Tab.~\ref{tab:channels}.  Similar reasoning holds for other channels.

\medskip

\begin{figure}
\begin{center}
  \includegraphics[width=80mm]{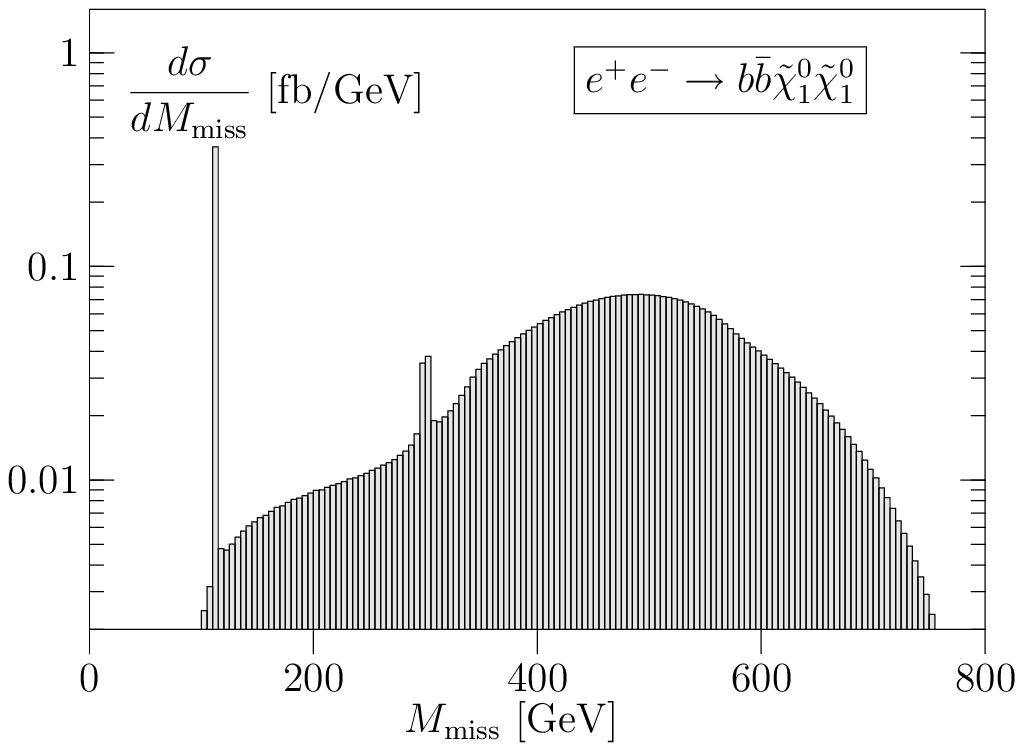}
  \includegraphics[width=80mm]{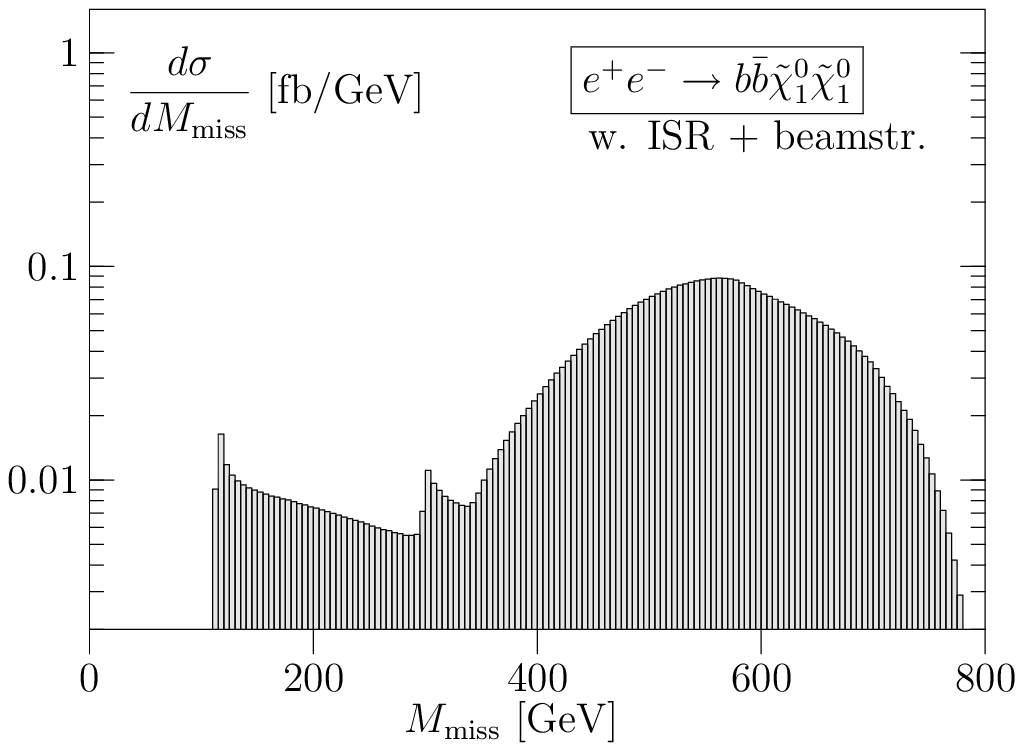}
\end{center}
\vspace*{-8mm}
\caption{Missing invariant mass spectrum for the full process
  $e^+e^-\to b\bar{b}\neua\neua$: on the left for the partonic
  process, on the right including ISR and beamstrahlung.}
\label{fig:mmiss}
\end{figure}

\medskip

The results in Tab.~\ref{tab:channels} also demonstrate that photon
radiation, both in the elementary process (ISR) and as a
semi-classical interaction of the incoming beams (beamstrahlung),
cannot be neglected.  For the numerical results, ISR is included using
the third-order leading-logarithmic approximation~\cite{dittmaier},
and beamstrahlung using the TESLA 800 parameterization in {\sc
circe}~\cite{circe}.  In both cases the photon radiation is
predominantly collinear with the incoming beams and therefore
invisible.  Therefore, all distributions depending on missing
momentum, \ie the momentum of the final-state neutralinos, are
distorted by such effects.  In the left panel of Fig.~\ref{fig:mmiss}
we show the missing invariant-mass spectrum for the full process
$e^+e^-\to b\bar{b}\neua\neua$ without ISR and beamstrahlung.  Two
narrow peaks are clearly visible, corresponding to the one light and
two (unresolved) heavy Higgs bosons.  These peaks sit on top of a
continuum reaching a maximum around $500\;\GeV$, dominantly stemming
from neutralino and sbottom pairs. We include ISR and beamstrahlung in
the right panel of Fig.~\ref{fig:mmiss}.  They tend to wash out the
two sharp peaks, with a long tail to higher invariant masses.  Without
explicitly showing it, we emphasize that the same happens to the SM
background, where the $Z$ boson decays invisibly into $\nu\bar\nu$.


\subsection{Isolating the sbottom-pair signal}

According to Tab.~\ref{tab:channels}, the dominant contribution to the
$b\bar{b}\neua\neua$ final state at an ILC is neutralino pair
production.  To study the sbottom sector, its contribution needs to be
isolated with kinematic cuts.  In addition, vector boson fusion into
$Z$ and Higgs bosons represent non-negligible backgrounds, and have to
be reduced accordingly.  We see that Higgs boson and heavy sbottom
production are of minor importance.

\begin{figure}
\begin{center}
\includegraphics[width=120mm]{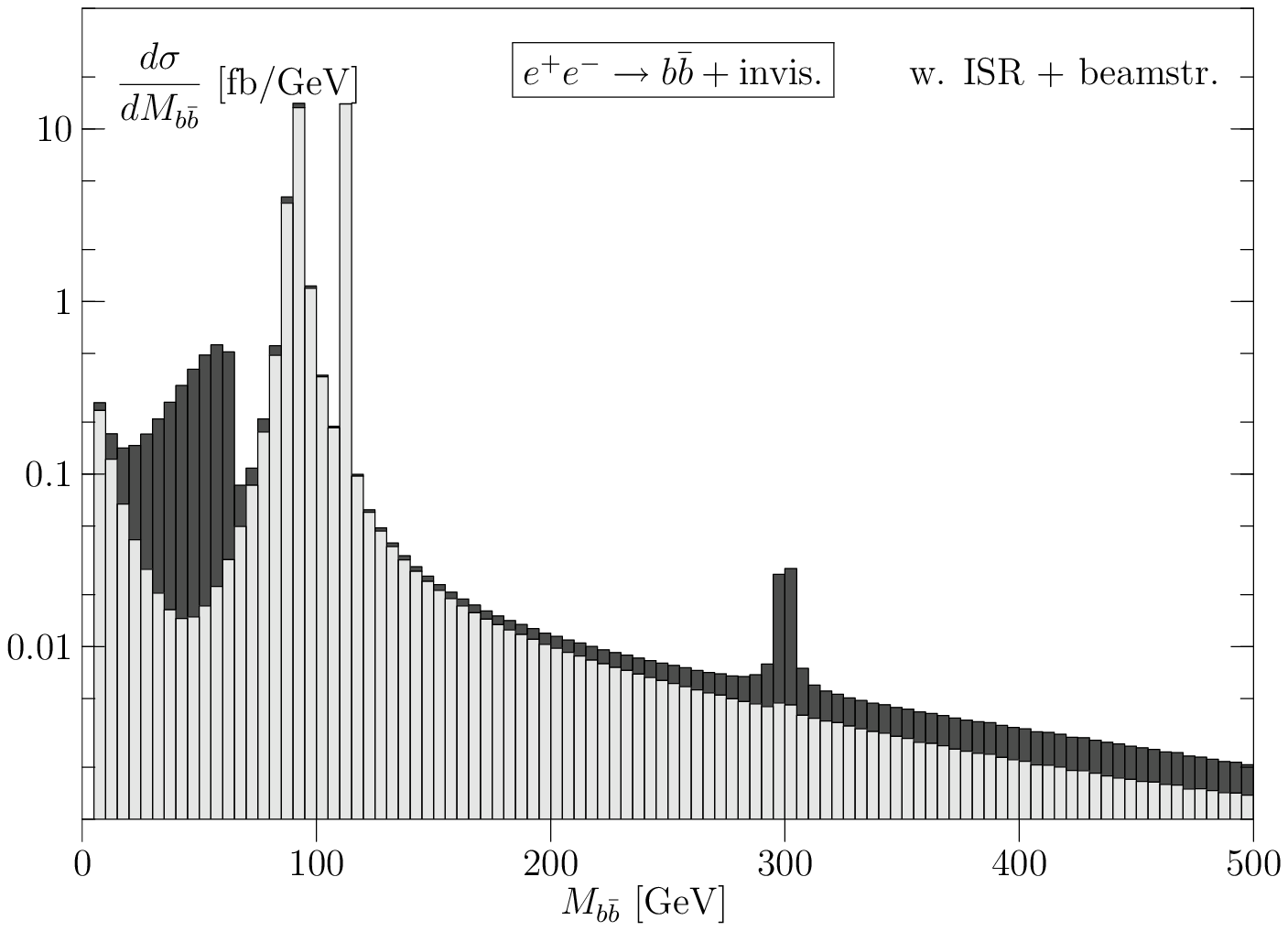}
\end{center}
\vspace*{-8mm}
\caption{The $b\bar{b}$ invariant mass spectrum for the full process 
  $e^+e^-\to b\bar{b}$+$\me$ with ISR and beamstrahlung.  The SM
  background ($Z\to\nu\bar{\nu}$) with the $Z,h$ peaks is light gray.
  Dark gray represents all MSSM processes, with two peaks from heavy
  neutralino and heavy Higgs decays.}
\label{fig:fullspectrum}
\end{figure}

An obvious cut for background reduction is on the reconstructed
$b\bar{b}$ invariant mass.  Fig.~\ref{fig:fullspectrum} shows the
distribution for the full process, with all Feynman diagrams and
including ISR and beamstrahlung.  SM contributions (light gray) and
the MSSM (dark) must be superimposed to obtain the complete signal and
background result, since neutrinos cannot be distinguished from
neutralinos.  The spectrum depicted in Fig.~\ref{fig:fullspectrum} has
several distinct features: there are narrow peaks at the $h$, $Z$ and
$H/A$ boson masses, as well as a broader enhancement around
$50\;\GeV$, associated with the $\neub$ three-body decay.  (The
$\sbl\sbl^*$ signal does not have any resonance structure and
populates the continuum at high invariant $b\bar{b}$ masses.)  To
remove all resonances we cut away the invariant mass windows:
\begin{equation}\label{cuts}
  M_{b\bar{b}} < 150\;\GeV \; , \qquad \qquad
  250\;\GeV < M_{b\bar{b}} < 350\;\GeV \; .
\end{equation}
This cut retains mostly sbottom-pair signal events, with some
continuum background.  In the crude NWA (just the simple production
channels $\sbl\sbl^*, \neua\neub, \neua\neuc$ and $W^+W^-\to Z/h$,
$ZZ$, $Zh$, $HA$, $\ldots$; times decay matrix elements), these cuts
would remove the entire background, while only marginally affecting
the signal.

\medskip

\begin{table}
\begin{displaymath}
  \begin{array}{|l|rr|}
    \hline
    \text{Channel} &  \sigma_{\rm BW}~[\fb] & \sigma_{\rm BW}^{\rm cut}~[\fb]
    \\
    \hline
    Zh           & 1.335 & 0.009
    \\
    ZH           & 0.000 & 0.000
    \\
    hA           & 0.000 & 0.000
    \\
    HA           & 0.314 & 0.003
    \\
    \neua\neub   &13.954 & 0.458
    \\
    \neua\neuc   & 4.828 & 0.454
    \\
    \neua\neud   & 0.938 & 0.937
    \\
    \sbl\sbl     & 0.757 & 0.451
    \\
    \sbl\sbr     & 0.002 & 0.001
    \\
    \hline
    \text{Sum}   &22.129 & 2.314
    \\
    \hline\hline
    \text{Exact} &19.624 & 0.487
    \\
    \text{w/ISR} &22.552 & 0.375
    \\
    \hline
  \end{array}
\hspace*{20mm}
  \begin{array}{|l|rr|}
    \hline
    \text{Channel} &  \sigma_{\rm BW}~[\fb] & \sigma_{\rm BW}^{\rm cut}~[\fb]
    \\
    \hline
    Z\bar\nu\nu  & 111.4 &  2.114
    \\
    h\bar\nu\nu  &  76.4 &  0.002
    \\
    H\bar\nu\nu  &   0.0 &  0.000
    \\
    \hline
    \text{Sum}   & 187.7 &  2.117
    \\
    \hline\hline
    \text{Exact} & 190.1 &  1.765
    \\
    \text{w/ISR} & 174.2 &  1.609
    \\
    \hline
  \end{array}
\end{displaymath}
\caption{SUSY cross sections contributing to 
  $e^+e^-\to b\bar{b}\neua\neua$ (left) and the SM background
  $e^+e^-\to b\bar{b}\nu\bar\nu$ (right).  The left column is the
  Breit-Wigner approximation without cuts.  The right column is after
  the $M_{b\bar{b}}$ cuts of Eq.(\protect\ref{cuts}). We show the
  results for the incoherent sum of channels, the complete result with
  all interferences, and the same with ISR and beamstrahlung.}
\label{tab:channels-cut}
\end{table}

We show the effect of applying this cut in Tab.~\ref{tab:channels-cut}
using the various approximations.  In the full calculation we retain
$60\%$ of the signal rate.  While in the on-shell approximation this
cut would remove $100\%$ of the peaked backgrounds, our complete
calculation including Breit-Wigner propagators retains a whopping
$2.3\,\fb$ (SUSY) and $2.1\,\fb$ (SM). 
Surprisingly, the exact tree-level cross section without ISR is
considerably smaller than that: $0.5\,\fb$ (SUSY, signal+background)
and $1.8\,\fb$ (SM).  Obviously, for the background SUSY processes the
Breit-Wigner approximation is misleadingly wrong if we force the phase
space into the sbottom-signal region.  Only the full calculation gives
a reliable result.

\smallskip

\begin{figure}
\includegraphics[width=85mm,height=75mm]{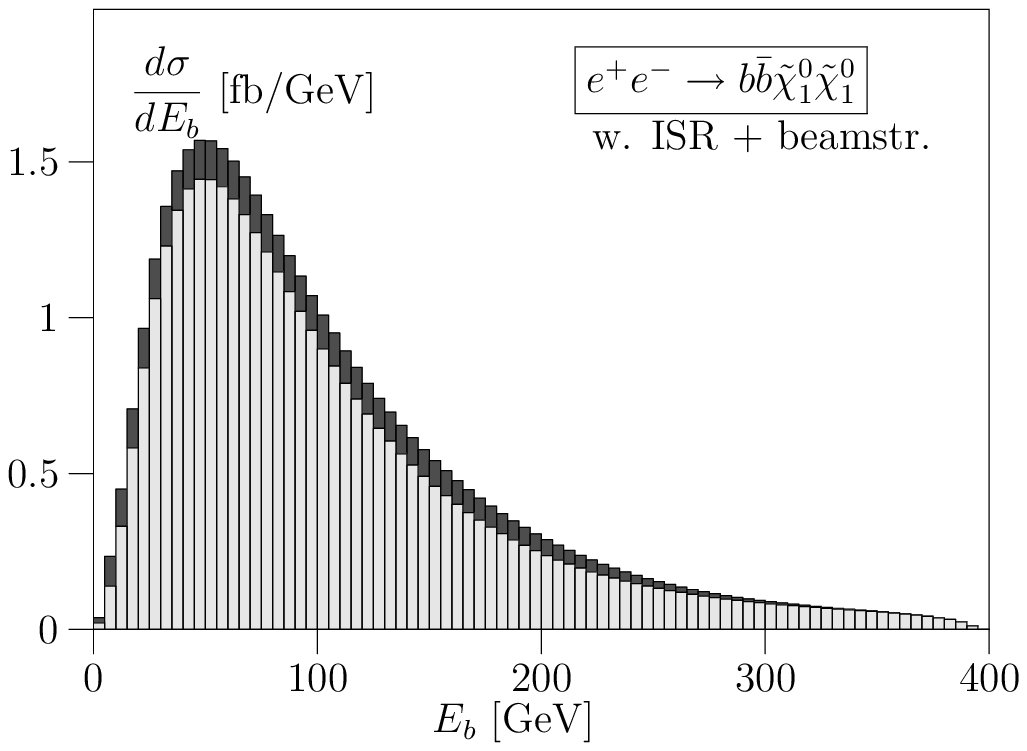}
\includegraphics[width=85mm,height=75mm]{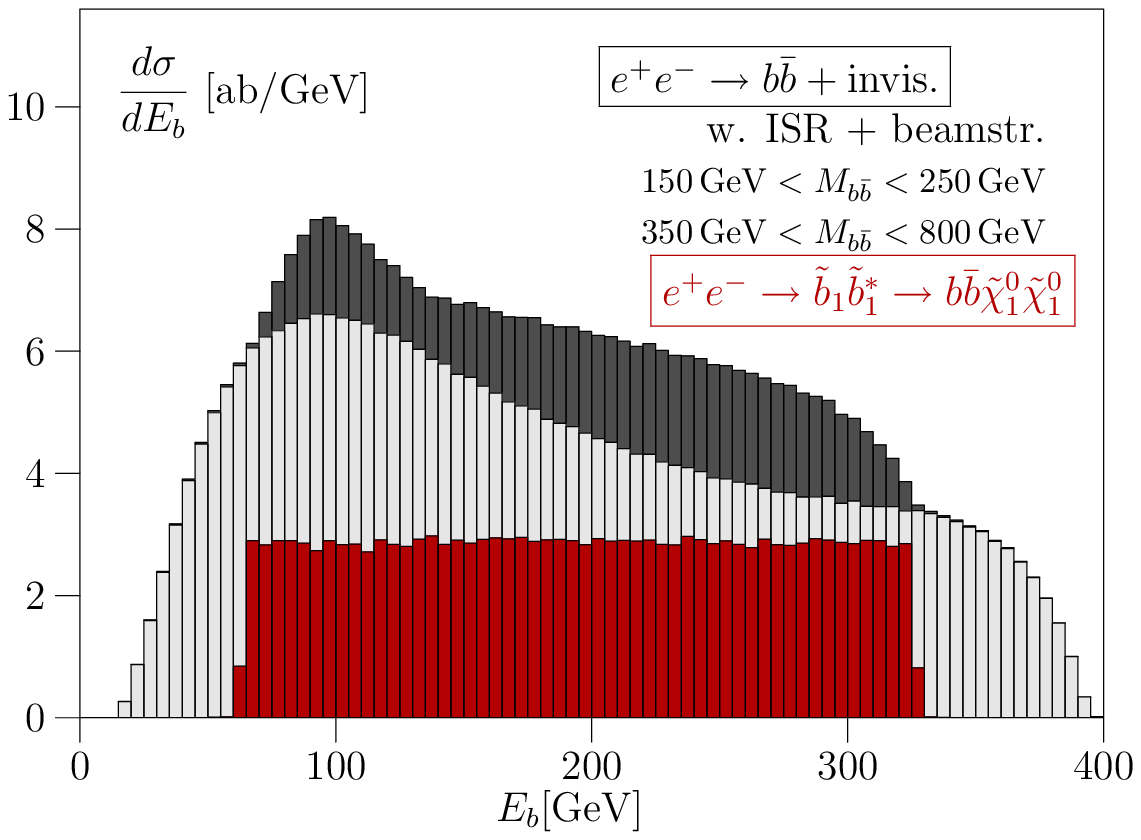}
\caption{The $E_b$ spectrum of the full process 
  $e^+e^-\to b\bar{b}$+$\me$, including all interferences and
  off-shell effects, plus ISR and beamstrahlung.  The light gray
  histogram is the SM background, dark gray the sum of SUSY processes.
  The left panel is before the cut of Eq.(\protect\ref{cuts}), while
  the right panel includes the cut.  Also in the right panel we show
  the idealized case (red) of on-shell sbottom production without ISR
  or beamstrahlung.  The SM background is again shown in light gray,
  while the dark gray shows the sbottom contribution alone.}
\label{fig:eb}
\end{figure}

In the absence of backgrounds, the $b$ jet energy spectrum from
sbottom decays exhibits a box-like shape corresponding to the decay
kinematics of $\sbl\to b\neua$.  Assuming that $m_{\neua}$ is known
from a threshold scan, the edges of the box would allow a simple
kinematical fit to yield a precise determination of $m_{\sbl}$.  The
realistic $E_b$ distribution appears in Fig.~\ref{fig:eb}.  In the
left panel we show the $E_b$ spectrum for the full process without
cuts, including all interferences, and taking ISR and beamstrahlung
into account.  The large background precludes any identification of a
box shape.  The right panel displays the same distribution after the
$M_{b\bar{b}}$ cuts of Eq.(\ref{cuts}) and compares it with the ideal
case (no background, no ISR, no cuts) in the same normalization.

The SUSY contribution after cuts (dark area) shows the same
kinematical limits as the ideal box, but the edges are washed out by
the combined effects of cuts, ISR/beamstrahlung, and continuum
background.  However, the signal sits atop a sizable leftover SM
background.  As argued above, this background cannot be realistically
simulated by simply concatenating particle production and decays.

\smallskip

Without going into detail, we note that for further improvement of the
signal-to-background ratio, one could use beam polarization (reducing
the $W^+W^-\to b\bar{b}$ continuum) or a cut on missing invariant mass
(to suppress $Z\to\nu\bar\nu$).  For a final verdict on the
measurement of sbottom properties in this decay channel, a realistic
analysis must also consider fragmentation, hadronization and detector
effects.  NLO corrections (at least; if not NNLO) to the signal
process must be taken into account to gain some idea about realistic
event rates.


\section{Summary and Outlook}

Phenomenological and experimental (Monte Carlo) analyses for new
physics at colliders are usually approached at a level of
sophistication which does not match the know-how we have from the
Standard Model.  For supersymmetric signals at the LHC and an ILC we
have carefully studied effects which occur beyond simple $2\to 2$
cross section analyses, using sbottom pair production as a simple
example process.

\medskip

At the LHC, the reconstruction of decay kinematics is the source of
essentially all information on heavy new particles.  Any observable
linked to cross sections instead of kinematical features is bound to
suffer from much larger QCD uncertainties.  Typical experimental
errors from jet energy scaling are of the same order as finite-width
effects in the total cross section.  However, in relevant
distributions, off-shell effects can easily be larger.

QCD off-shell effects also include additional jet radiation from the
incoming state.  Usually, jet radiation is treated by parton showers
in the collinear approximation.  For processes with bottom jets in the
final state we tested this approximation by computing the effects of
two additional bottom jets created through gluon splitting in the
initial state.  The effects on the rate are typically below $10\%$,
and kinematical distributions do indeed change.  In our case,
distinguishing between initial-state bottom jets and decay bottom jets
via rapidity and transverse momentum characteristics does not look
promising.

Sbottom pair production at the LHC has the fortunate feature that most
of these off-shell effects and combinatorial backgrounds can be
removed together with the SM backgrounds, but this feature is by no
means guaranteed for general SUSY processes.

\medskip

At an ILC, the extraction of parameters from kinematic distributions
is usually more precise compared to more inclusive measurements.  In
contrast to the LHC, the typical size $\Gamma/M$ of off-shell effects
exceeds the present ILC design experimental precision.  It is
therefore mandatory for multi-particle final states to include the
complete set of off-shell Feynman diagrams in ILC studies, since they
can alter signal distributions drastically.  This was impressively
demonstrated by our study of sbottom pair production where we found up
to $400\%$ corrections to production rates from off-shell effects,
after standard cuts.

Irreducible SM backgrounds to missing energy signals can strongly
distort the shapes of energy and invariant mass distributions.  Hence,
if we would attempt to extract masses and mass differences from
invariant mass distributions at an ILC, we find that we must take into
account off-shell effects and additional many-particle intermediate
states which can change cross sections dramatically.  Simulation of
initial state radiation and beamstrahlung is mandatory to describe
shapes of resonances and distributions in a realistic linear collider
environment.

\medskip

To compute the effects described above we implemented the MSSM
Lagrangian and the proper description of Majorana particles in the
matrix element generators {\sc madgraph}/{\sc mad\-event}, {\sc
o'mega}/{\sc whizard} and {\sc amegic}\texttt{++}/{\sc sherpa}.  To
carefully check these extensions we compared several hundred SUSY
production processes numerically, as well as performed a number of
unitarity and gauge invariance checks.  All results, as well as the
SLHA input file, are given in the Appendix --- we are confident that
this list of processes can serve as a standard reference to generally
check MSSM implementations in collider physics or phenomenology tools.


\subsection*{Acknowledgments}

All members of this long-term collaboration on the
Comparison of
Automated 
Tools for
Phenomenological
Investigations of
SuperSymmetry
are grateful to the DESY Theory Group, the Madison Pheno Institute,
the Institute for Theoretical Physics at Dresden, and MPI Munich for
their continuous support and hospitality.  We would also like to thank
the Aspen Center for Physics, where much of this work was completed.
We would like to thank Gudrun Hiller for her help with the flavor
constraints.  K.~H. is supported in part by Grant-in-Aid for
Scientific Research of MEXT, Japan, No. 17540281.  T.~O., J.~R. and
W.~K. are supported in part by the German Helmholtz-Gemeinschaft,
Grant No.\ VH--NG--005.  T.~O. is supported in part by
Bundesministerium f\"ur Bildung und Forschung Germany, Grant No.\
05HT1WWA/2.  D.R. is supported in part by the U.S. Department of
Energy under grant No.\ DE-FG02-91ER40685.  F.K. and S.S. gratefully
acknowledge financial support by BMBF.


\appendix 

\clearpage


\section{Input Parameters Used in the Comparison}
\label{sec:input}

Here we list the input parameters we used, which are in the blocks
relevant for our purposes from the SLHA output of the
{\sc softsusy} program:

\begin{quote}\footnotesize
\begin{verbatim}
# SOFTSUSY1.9
# B.C. Allanach, Comput. Phys. Commun. 143 (2002) 305-331, hep-ph/0104145
Block SPINFO         # Program information
     1   SOFTSUSY    # spectrum calculator
     2   1.9         # version number
Block MODSEL  # Select model
     1    1   # sugra
Block SMINPUTS   # Standard Model inputs
#     1    1.27934000e+02   # alpha_em^(-1)(MZ) SM MSbar
     2    1.16639000e-05   # G_Fermi
#     3    1.17200000e-01   # alpha_s(MZ)MSbar
#     4    9.11876000e+01   # MZ(pole)
#     5    4.25000000e+00   # Mb(mb)
#     6    1.74300000e+02   # Mtop(pole)
     7    1.77700000e+00   # Mtau(pole)
Block MINPAR  # SUSY breaking input parameters
     3    1.00000000e+01   # tanb
     4    1.00000000e+00   # sign(mu)
     1    1.00000000e+02   # m0
     2    2.50000000e+02   # m12
     5   -1.00000000e+02   # A0
# Low energy data in SOFTSUSY: MIXING=-1 TOLERANCE=1.00000000e-03
# mgut=2.46245508e+16 GeV
Block MASS   # Mass spectrum
#PDG code      mass              particle
        24     8.04194155e+01   # MW
        25     1.10762900e+02   # h0
        35     4.00615086e+02   # H0
        36     4.00247030e+02   # A0
        37     4.08528577e+02   # H+
   1000001     5.72715810e+02   # ~d_L
   1000002     5.67266777e+02   # ~u_L
   1000003     5.72715810e+02   # ~s_L
   1000004     5.67266777e+02   # ~c_L
   1000005     5.15224253e+02   # ~b_1
   1000006     3.95930570e+02   # ~t_1
   1000011     2.04280587e+02   # ~e_L
   1000012     1.88661921e+02   # ~nue_L
   1000013     2.04280587e+02   # ~mu_L
   1000014     1.88661921e+02   # ~numu_L
   1000015     1.36227332e+02   # ~stau_1
   1000016     1.87777460e+02   # ~nu_tau_L
   1000021     6.07618238e+02   # ~g
   1000022     9.72807171e+01   # ~neutralino(1)
   1000023     1.80959888e+02   # ~neutralino(2)
   1000024     1.80377023e+02   # ~chargino(1)
   1000025    -3.64450624e+02   # ~neutralino(3)
   1000035     3.83149239e+02   # ~neutralino(4)
   1000037     3.83385634e+02   # ~chargino(2)
   2000001     5.46084642e+02   # ~d_R
   2000002     5.47013902e+02   # ~u_R
   2000003     5.46084642e+02   # ~s_R
   2000004     5.47013902e+02   # ~c_R
   2000005     5.43980537e+02   # ~b_2
   2000006     5.85709387e+02   # ~t_2
   2000011     1.45527209e+02   # ~e_R
   2000013     1.45527209e+02   # ~mu_R
   2000015     2.08226705e+02   # ~stau_2
# Higgs mixing
Block alpha   # Effective Higgs mixing parameter
          -1.13731924e-01   # alpha
Block stopmix  # stop mixing matrix
   1  1     5.38076009e-01   # O_{11}
   1  2     8.42896322e-01   # O_{12}
   2  1     8.42896322e-01   # O_{21}
   2  2    -5.38076009e-01   # O_{22}
Block sbotmix  # sbottom mixing matrix
   1  1     9.47748557e-01   # O_{11}
   1  2     3.19018296e-01   # O_{12}
   2  1    -3.19018296e-01   # O_{21}
   2  2     9.47748557e-01   # O_{22}
Block staumix  # stau mixing matrix
  1  1     2.80949722e-01   # O_{11}
  1  2     9.59722488e-01   # O_{12}
  2  1     9.59722488e-01   # O_{21}
  2  2    -2.80949722e-01   # O_{22}
Block nmix  # neutralino mixing matrix
  1  1     9.86069014e-01   # N_{1,1}
  1  2    -5.46217310e-02   # N_{1,2}
  1  3     1.47637908e-01   # N_{1,3}
  1  4    -5.37346696e-02   # N_{1,4}
  2  1     1.02047560e-01   # N_{2,1}
  2  2     9.42730347e-01   # N_{2,2}
  2  3    -2.74969181e-01   # N_{2,3}
  2  4     1.58863895e-01   # N_{2,4}
  3  1    -6.04553550e-02   # N_{3,1}
  3  2     8.97014273e-02   # N_{3,2}
  3  3     6.95501771e-01   # N_{3,3}
  3  4     7.10335196e-01   # N_{3,4}
  4  1    -1.16616232e-01   # N_{4,1}
  4  2     3.16590608e-01   # N_{4,2}
  4  3     6.47203433e-01   # N_{4,3}
  4  4    -6.83592537e-01   # N_{4,4}
Block Umix  # chargino U mixing matrix 
  1  1     9.15543496e-01   # U_{1,1}
  1  2    -4.02218978e-01   # U_{1,2}
  2  1     4.02218978e-01   # U_{2,1}
  2  2     9.15543496e-01   # U_{2,2}
Block Vmix  # chargino V mixing matrix 
  1  1     9.72352114e-01   # V_{1,1}
  1  2    -2.33519522e-01   # V_{1,2}
  2  1     2.33519522e-01   # V_{2,1}
  2  2     9.72352114e-01   # V_{2,2}
Block hmix Q= 4.64241862e+02  # Higgs mixing parameters
     1     3.58355327e+02   # mu(Q)MSSM DRbar
#     2     9.75144517e+00   # tan beta(Q)MSSM DRbar
     3     2.44921676e+02   # higgs vev(Q)MSSM DRbar
     4     1.69588951e+04   # mA^2(Q)MSSM DRbar
Block au Q= 4.64241862e+02  
  1  1     0.00000000e+00   # Au(Q)MSSM DRbar
  2  2     0.00000000e+00   # Ac(Q)MSSM DRbar
  3  3    -5.04528807e+02   # At(Q)MSSM DRbar
Block ad Q= 4.64241862e+02  
  1  1     0.00000000e+00   # Ad(Q)MSSM DRbar
  2  2     0.00000000e+00   # As(Q)MSSM DRbar
  3  3    -7.97132778e+02   # Ab(Q)MSSM DRbar
Block ae Q= 4.64241862e+02  
  1  1     0.00000000e+00   # Ae(Q)MSSM DRbar
  2  2     0.00000000e+00   # Amu(Q)MSSM DRbar
  3  3    -2.56155534e+02   # Atau(Q)MSSM DRbar

\end{verbatim}
\end{quote}

Parameters used with a different value than specified in the above
SLHA file are $M_W=80.419\,\GeV$, $M_Z=91.188\,\GeV$.  We set all SUSY
particle widths to zero, since there are no SUSY particles in the
$s$-channel.  (The spectrum generator {\sc softsusy} does not
calculate the widths of the SUSY particles.  This is instead done by
the program {\sc sdecay}~\cite{sdecay}.)  The only widths used in our
comparison are set by hand, $\Gamma_W=2.048\,\GeV$ and
$\Gamma_Z=2.446\,\GeV$.  All Higgs widths have been set to zero, as
well as the electron mass.  The third generation quark masses have
been given the values $m_t=178.0\,\GeV$ and $m_b=4.6\,\GeV$. For the
strong coupling we take $\alpha_s(M_Z)=0.118$.  The $G_F-M_Z-\alpha$
scheme has been used for the SM parameters.



\bigskip\bigskip

\section{Cross Section Values for \boldmath$2\to 2$ SUSY Processes}
\label{sec:xsec}

The following tables are also maintained at the web page \\
\url{http://www.sherpa-mc.de/susy_comparison/susy_comparison.html}.

\subsection{\boldmath$e^+e^-$ processes}

  \begin{footnotesize}
\begin{tabular}{|c||ll|ll|ll|}\hline
    \multicolumn{7}{|c|}{$\bf{e^+e^-\to X}$} \\\hline
    Final &  \multicolumn{2}{c|}{{\sc madgraph}/{\sc helas}} & 
    \multicolumn{2}{c|}{{\sc o'mega}/{\sc whizard}} & \multicolumn{2}{c|}{{\sc amegic++/sherpa}}\\
    state
    & \multicolumn{1}{c}{0.5 TeV} 
    & \multicolumn{1}{c|}{2 TeV} 
    & \multicolumn{1}{c}{0.5 TeV} 
    & \multicolumn{1}{c|}{2 TeV} 
    & \multicolumn{1}{c}{0.5 TeV} 
    & \multicolumn{1}{c|}{2 TeV} 
    \\
\hline
  \end{tabular}\label{xsec:e+e-}
  \end{footnotesize}

\clearpage
\subsection{\boldmath$\tau^+\tau^-$ processes}

  \begin{footnotesize}
  \begin{tabular}{|c||ll|ll|ll|}\hline
    \multicolumn{7}{|c|}{$\bf{\tau^+\tau^-\to X}$} \\\hline
    Final &  \multicolumn{2}{c|}{{\sc madgraph}/{\sc helas}} & 
    \multicolumn{2}{c|}{{\sc o'mega}/{\sc whizard}} & \multicolumn{2}{c|}{{\sc amegic++/sherpa}}\\
    state
    & \multicolumn{1}{c}{0.5 TeV} 
    & \multicolumn{1}{c|}{2 TeV} 
    & \multicolumn{1}{c}{0.5 TeV} 
    & \multicolumn{1}{c|}{2 TeV} 
    & \multicolumn{1}{c}{0.5 TeV} 
    & \multicolumn{1}{c|}{2 TeV} 
    \\
\hline
  \end{tabular}\label{xsec:tau+tau-}
  \end{footnotesize}

\clearpage
\subsection{\boldmath$e^-\bar\nu_e$ processes}

  \begin{footnotesize}
  \begin{tabular}{|c||ll|ll|ll|}\hline
    \multicolumn{7}{|c|}{$\bf{e^-\bar\nu_e\to X^-}$} \\\hline
    Final &  \multicolumn{2}{c|}{{\sc madgraph}/{\sc helas}} & 
    \multicolumn{2}{c|}{{\sc o'mega}/{\sc whizard}} & \multicolumn{2}{c|}{{\sc amegic++/sherpa}}\\
    state
    & \multicolumn{1}{c}{0.5 TeV} 
    & \multicolumn{1}{c|}{2 TeV} 
    & \multicolumn{1}{c}{0.5 TeV} 
    & \multicolumn{1}{c|}{2 TeV} 
    & \multicolumn{1}{c}{0.5 TeV} 
    & \multicolumn{1}{c|}{2 TeV} 
    \\\hline\hline 
\tabc{\sel\sne^*}{158.69(1)}{\pone67.096(5)}{158.694(7)}{\pone67.095(3)}{158.703(8)}{\pone67.100(3)}
\tabc{\ser\sne^*}{\pone68.51(1)}{\pten6.547(4)}{\pone68.513(3)}{\pten6.5470(4)}{\pone68.508(3)}{\pten6.5469(3)}
\tabc{\sml\snm^*}{\pone58.492(3)}{\pone13.894(1)}{\pone58.491(2)}{\pone13.8935(5)}{\pone58.492(3)}{\pone13.8931(7)}
\tabc{\staul\snt^*}{\pten8.5018(5)}{\pten1.1169(1)}{\pten8.5021(3)}{\pten1.11690(4)}{\pten8.5018(4)}{\pten1.11696(6)}
\tabc{\staur\snt^*}{\pone51.792(3)}{\pone12.784(1)}{\pone51.790(2)}{\pone12.7836(5)}{\pone51.795(3)}{\pone12.7844(6)}
\hline
\tabc{\cham\neua}{137.414(5)}{\pone21.4202(6)}{137.416(4)}{\pone21.4203(9)}{137.426(8)}{\pone21.419(2)}
\tabc{\cham\neub}{\pone58.797(3)}{\pone21.284(2)}{\pone58.795(1)}{\pone21.283(1)}{\pone58.794(3)}{\pone21.282(2)}
\tabc{\cham\neuc}{\mcent}{\pten2.2676(1)}{\mcent}{\pten2.26760(7)}{\mcent}{\pten2.2678(1)}
\tabc{\cham\neud}{\mcent}{\pten3.5104(2)}{\mcent}{\pten3.51046(6)}{\mcent}{\pten3.5105(2)}
\tabc{\chbm\neua}{\pten1.16070(5)}{\pten1.73602(6)}{\pten1.16072(3)}{\pten1.73607(6)}{\pten1.16066(6)}{\pten1.73593(9)}
\tabc{\chbm\neub}{\mcent}{\pten3.6111(3)}{\mcent}{\pten3.61122(6)}{\mcent}{\pten3.6113(2)}
\tabc{\chbm\neuc}{\mcent}{\pone26.9497(5)}{\mcent}{\pone26.9511(7)}{\mcent}{\pone26.952(1)}
\tabc{\chbm\neud}{\mcent}{\pone24.022(1)}{\mcent}{\pone24.0223(8)}{\mcent}{\pone24.022(1)}
\hline
  \end{tabular}\label{xsec:e-nu}
  \end{footnotesize}


\subsection{\boldmath$\tau^-\bar\nu_\tau$ processes}

  \begin{footnotesize}
  \begin{tabular}{|c||ll|ll|ll|}\hline
    \multicolumn{7}{|c|}{$\bf{\tau^-\bar\nu_\tau\to X^-}$} \\\hline
    Final &  \multicolumn{2}{c|}{{\sc madgraph}/{\sc helas}} & 
    \multicolumn{2}{c|}{{\sc o'mega}/{\sc whizard}} & \multicolumn{2}{c|}{{\sc amegic++/sherpa}}\\
    state
    & \multicolumn{1}{c}{0.5 TeV} 
    & \multicolumn{1}{c|}{2 TeV} 
    & \multicolumn{1}{c}{0.5 TeV} 
    & \multicolumn{1}{c|}{2 TeV} 
    & \multicolumn{1}{c}{0.5 TeV} 
    & \multicolumn{1}{c|}{2 TeV} 
    \\
\hline
  \end{tabular}\label{xsec:tau-nu}
  \end{footnotesize}


\subsection{\boldmath$b\bar{t}$ processes}

  \begin{footnotesize}
  \begin{tabular}{|c||ll|ll|ll|}\hline
    \multicolumn{7}{|c|}{$\bf{b\bar{t}\to X^-}$} \\\hline
    Final &  \multicolumn{2}{c|}{{\sc madgraph}/{\sc helas}} & 
    \multicolumn{2}{c|}{{\sc o'mega}/{\sc whizard}} & \multicolumn{2}{c|}{{\sc amegic++/sherpa}}\\
    state
    & \multicolumn{1}{c}{0.5 TeV} 
    & \multicolumn{1}{c|}{2 TeV} 
    & \multicolumn{1}{c}{0.5 TeV} 
    & \multicolumn{1}{c|}{2 TeV} 
    & \multicolumn{1}{c}{0.5 TeV} 
    & \multicolumn{1}{c|}{2 TeV} 
    \\
\hline
  \end{tabular}\label{xsec:bt}
  \end{footnotesize}

\clearpage
\subsection{\boldmath$W^+W^-$ processes}

  \begin{footnotesize}
  \begin{tabular}{|c||ll|ll|ll|}\hline
    \multicolumn{7}{|c|}{$\bf{W^+W^-\to X}$} \\\hline
    Final &  \multicolumn{2}{c|}{{\sc madgraph}/{\sc helas}} & 
    \multicolumn{2}{c|}{{\sc o'mega}/{\sc whizard}} & \multicolumn{2}{c|}{{\sc amegic++/sherpa}}\\
    state
    & \multicolumn{1}{c}{0.5 TeV} 
    & \multicolumn{1}{c|}{2 TeV} 
    & \multicolumn{1}{c}{0.5 TeV} 
    & \multicolumn{1}{c|}{2 TeV} 
    & \multicolumn{1}{c}{0.5 TeV} 
    & \multicolumn{1}{c|}{2 TeV} 
    \\
\hline
  \end{tabular}\label{xsec:W+W-}
  \end{footnotesize}

\clearpage
\subsection{\boldmath$W^-Z$ processes}

  \begin{footnotesize}
  \begin{tabular}{|c||ll|ll|ll|}\hline
    \multicolumn{7}{|c|}{$\bf{W^-Z\to X^-}$} \\\hline
    Final &  \multicolumn{2}{c|}{{\sc madgraph}/{\sc helas}} & 
    \multicolumn{2}{c|}{{\sc o'mega}/{\sc whizard}} & \multicolumn{2}{c|}{{\sc amegic++/sherpa}}\\
    state
    & \multicolumn{1}{c}{0.5 TeV} 
    & \multicolumn{1}{c|}{2 TeV} 
    & \multicolumn{1}{c}{0.5 TeV} 
    & \multicolumn{1}{c|}{2 TeV} 
    & \multicolumn{1}{c}{0.5 TeV} 
    & \multicolumn{1}{c|}{2 TeV} 
    \\
\hline
 \end{tabular}\label{xsec:W-Z}
  \end{footnotesize}


\subsection{\boldmath$W^-\gamma$ processes}

  \begin{footnotesize}
  \begin{tabular}{|c||ll|ll|ll|}\hline
    \multicolumn{7}{|c|}{$\bf{W^-\gamma\to X^-}$} \\\hline
    Final &  \multicolumn{2}{c|}{{\sc madgraph}/{\sc helas}} & 
    \multicolumn{2}{c|}{{\sc o'mega}/{\sc whizard}} & \multicolumn{2}{c|}{{\sc amegic++/sherpa}}\\
    state
    & \multicolumn{1}{c}{0.5 TeV} 
    & \multicolumn{1}{c|}{2 TeV} 
    & \multicolumn{1}{c}{0.5 TeV} 
    & \multicolumn{1}{c|}{2 TeV} 
    & \multicolumn{1}{c}{0.5 TeV} 
    & \multicolumn{1}{c|}{2 TeV} 
    \\
\hline
 \end{tabular}\label{xsec:W-A}
  \end{footnotesize}

\clearpage
\subsection{\boldmath$ZZ$ processes}

  \begin{footnotesize}
  \begin{tabular}{|c||ll|ll|ll|}\hline
    \multicolumn{7}{|c|}{$\bf{ZZ\to X}$} \\\hline
    Final &  \multicolumn{2}{c|}{{\sc madgraph}/{\sc helas}} & 
    \multicolumn{2}{c|}{{\sc o'mega}/{\sc whizard}} & \multicolumn{2}{c|}{{\sc amegic++/sherpa}}\\
    state
    & \multicolumn{1}{c}{0.5 TeV} 
    & \multicolumn{1}{c|}{2 TeV} 
    & \multicolumn{1}{c}{0.5 TeV} 
    & \multicolumn{1}{c|}{2 TeV} 
    & \multicolumn{1}{c}{0.5 TeV} 
    & \multicolumn{1}{c|}{2 TeV} 
    \\
\hline
  \end{tabular}\label{xsec:ZZ}
  \end{footnotesize}
\clearpage
\subsection{\boldmath$Z\gamma$ processes}

  \begin{footnotesize}
  \begin{tabular}{|c||ll|ll|ll|}\hline
    \multicolumn{7}{|c|}{$\bf{Z\gamma\to X}$} \\\hline
    Final &  \multicolumn{2}{c|}{{\sc madgraph}/{\sc helas}} & 
    \multicolumn{2}{c|}{{\sc o'mega}/{\sc whizard}} & \multicolumn{2}{c|}{{\sc amegic++/sherpa}}\\
    state
    & \multicolumn{1}{c}{0.5 TeV} 
    & \multicolumn{1}{c|}{2 TeV} 
    & \multicolumn{1}{c}{0.5 TeV} 
    & \multicolumn{1}{c|}{2 TeV} 
    & \multicolumn{1}{c}{0.5 TeV} 
    & \multicolumn{1}{c|}{2 TeV} 
    \\
\hline
  \end{tabular}\label{xsec:ZA}
  \end{footnotesize}

\clearpage
\subsection{\boldmath$\gamma\gamma$ processes}

  \begin{footnotesize}
  \begin{tabular}{|c||ll|ll|ll|}\hline
    \multicolumn{7}{|c|}{$\bf{\gamma\gamma\to X}$} \\\hline
    Final &  \multicolumn{2}{c|}{{\sc madgraph}/{\sc helas}} & 
    \multicolumn{2}{c|}{{\sc o'mega}/{\sc whizard}} & \multicolumn{2}{c|}{{\sc amegic++/sherpa}}\\
    state
    & \multicolumn{1}{c}{0.5 TeV} 
    & \multicolumn{1}{c|}{2 TeV} 
    & \multicolumn{1}{c}{0.5 TeV} 
    & \multicolumn{1}{c|}{2 TeV} 
    & \multicolumn{1}{c}{0.5 TeV} 
    & \multicolumn{1}{c|}{2 TeV} 
    \\
\hline
 \end{tabular}\label{xsec:AA}
  \end{footnotesize}


\subsection{\boldmath$g\gamma$ processes}

  \begin{footnotesize}
  \begin{tabular}{|c||ll|ll|ll|}\hline
    \multicolumn{7}{|c|}{$\bf{g\gamma\to X}$} \\\hline
    Final &  \multicolumn{2}{c|}{{\sc madgraph}/{\sc helas}} & 
    \multicolumn{2}{c|}{{\sc o'mega}/{\sc whizard}} & \multicolumn{2}{c|}{{\sc amegic++/sherpa}}\\
    state
    & \multicolumn{1}{c}{0.5 TeV} 
    & \multicolumn{1}{c|}{2 TeV} 
    & \multicolumn{1}{c}{0.5 TeV} 
    & \multicolumn{1}{c|}{2 TeV} 
    & \multicolumn{1}{c}{0.5 TeV} 
    & \multicolumn{1}{c|}{2 TeV} 
    \\\hline\hline 
\tabc{\sul\sul^*}{\mcent}{55.427(1)}{\mcent}{55.4290(8)}{\mcent}{55.428(3)}
\tabc{\sur\sur^*}{\mcent}{57.017(1)}{\mcent}{57.0184(9)}{\mcent}{57.020(3)}
\tabc{\scl\scl^*}{\mcent}{55.427(1)}{\mcent}{55.4288(8)}{\mcent}{55.430(3)}
\tabc{\scr\scr^*}{\mcent}{57.017(1)}{\mcent}{57.0175(9)}{\mcent}{57.019(3)}
\tabc{\stl\stl^*}{\mcent}{73.382(2)}{\mcent}{73.382(2)}{\mcent}{73.383(4)}
\tabc{\str\str^*}{\mcent}{54.113(1)}{\mcent}{54.1136(8)}{\mcent}{54.113(3)}
\hline
\tabc{\sdl\sdl^*}{\mcent}{13.7565(4)}{\mcent}{13.7569(2)}{\mcent}{13.7560(7)}
\tabc{\sdr\sdr^*}{\mcent}{14.2733(4)}{\mcent}{14.2737(2)}{\mcent}{14.2740(7)}
\tabc{\ssl\ssl^*}{\mcent}{13.7565(4)}{\mcent}{13.7568(2)}{\mcent}{13.7575(7)}
\tabc{\ssr\ssr^*}{\mcent}{14.2733(4)}{\mcent}{14.2735(2)}{\mcent}{14.2731(7)}
\tabc{\sbl\sbl^*}{\mcent}{14.9542(4)}{\mcent}{14.9546(3)}{\mcent}{14.9540(7)}
\tabc{\sbr\sbr^*}{\mcent}{14.3169(4)}{\mcent}{14.3171(2)}{\mcent}{14.3183(7)}
\hline
 \end{tabular}\label{xsec:GA}
  \end{footnotesize}


\subsection{\boldmath$gZ$ processes}

  \begin{footnotesize}
  \begin{tabular}{|c||ll|ll|ll|}\hline
    \multicolumn{7}{|c|}{$\bf{gZ \to X}$} \\\hline
    Final &  \multicolumn{2}{c|}{{\sc madgraph}/{\sc helas}} & 
    \multicolumn{2}{c|}{{\sc o'mega}/{\sc whizard}} & \multicolumn{2}{c|}{{\sc amegic++/sherpa}}\\
    state 
    & \multicolumn{1}{c}{0.5 TeV} 
    & \multicolumn{1}{c|}{2 TeV} 
    & \multicolumn{1}{c}{0.5 TeV} 
    & \multicolumn{1}{c|}{2 TeV} 
    & \multicolumn{1}{c}{0.5 TeV} 
    & \multicolumn{1}{c|}{2 TeV} 
    \\\hline\hline 
\tabc{\sul\sul^*}{\mcent}{\pone59.776(2)}{\mcent}{\pone59.7774(9)}{\mcent}{\pone59.778(3)}
\tabc{\sur\sur^*}{\mcent}{\pone10.9082(3)}{\mcent}{\pone10.9085(2)}{\mcent}{\pone10.9079(5)}
\tabc{\scl\scl^*}{\mcent}{\pone59.776(2)}{\mcent}{\pone59.7772(9)}{\mcent}{\pone59.778(3)}
\tabc{\scr\scr^*}{\mcent}{\pone10.9082(3)}{\mcent}{\pone10.9084(2)}{\mcent}{\pone10.9088(5)}
\tabc{\stl\stl^*}{\mcent}{\pten0.0074179(2)}{\mcent}{\pten0.0074179(2)}{\mcent}{\pten0.0074182(4)}
\tabc{\str\str^*}{\mcent}{\pone20.2088(5)}{\mcent}{\pone20.2091(3)}{\mcent}{\pone20.208(1)}
\tabc{\stl\str^*}{\mcent}{111.978(3)}{\mcent}{111.986(4)}{\mcent}{111.980(6)}
\hline
\tabc{\sdl\sdl^*}{\mcent}{\pone86.956(2)}{\mcent}{\pone86.9615(1)}{\mcent}{\pone86.960(4)}
\tabc{\sdr\sdr^*}{\mcent}{\pten2.73075(7)}{\mcent}{\pten2.73090(4)}{\mcent}{\pten2.7308(1)}
\tabc{\ssl\ssl^*}{\mcent}{\pone86.956(2)}{\mcent}{\pone86.959(1)}{\mcent}{\pone86.956(4)}
\tabc{\ssr\ssr^*}{\mcent}{\pten2.73075(7)}{\mcent}{\pten2.73078(4)}{\mcent}{\pten2.7308(1)}
\tabc{\sbl\sbl^*}{\mcent}{\pone73.359(2)}{\mcent}{\pone73.360(1)}{\mcent}{\pone73.354(4)}
\tabc{\sbr\sbr^*}{\mcent}{\pten0.268544(7)}{\mcent}{\pten0.268554(4)}{\mcent}{\pten0.26857(1)}
\tabc{\sbl\sbr^*}{\mcent}{\pone12.5213(3)}{\mcent}{\pone12.5214(2)}{\mcent}{\pone12.5214(6)}
\hline
 \end{tabular}\label{xsec:GZ}
  \end{footnotesize}


\subsection{\boldmath$gW^-$ processes}

  \begin{footnotesize}
  \begin{tabular}{|c||ll|ll|ll|}\hline
    \multicolumn{7}{|c|}{$\bf{gW^-\to X^-}$} \\\hline
    Final &  \multicolumn{2}{c|}{{\sc madgraph}/{\sc helas}} & 
    \multicolumn{2}{c|}{{\sc o'mega}/{\sc whizard}} & \multicolumn{2}{c|}{{\sc amegic++/sherpa}}\\
    state
    & \multicolumn{1}{c}{0.5 TeV} 
    & \multicolumn{1}{c|}{2 TeV} 
    & \multicolumn{1}{c}{0.5 TeV} 
    & \multicolumn{1}{c|}{2 TeV} 
    & \multicolumn{1}{c}{0.5 TeV} 
    & \multicolumn{1}{c|}{2 TeV} 
    \\\hline\hline 
\tabc{\sdl\sul^*}{\mcent}{187.611(5)}{\mcent}{187.616(3)}{\mcent}{187.604(8)}
\tabc{\ssr\scr^*}{\mcent}{187.611(5)}{\mcent}{187.617(3)}{\mcent}{187.619(8)}
\tabc{\sbl\stl^*}{\mcent}{138.625(2)}{\mcent}{138.625(4)}{\mcent}{138.624(7)}
\tabc{\sbr\str^*}{\mcent}{\pone16.5094(3)}{\mcent}{\pone16.5095(3)}{\mcent}{\pone16.5088(8)}
\tabc{\sbl\str^*}{\mcent}{195.686(1)}{\mcent}{195.692(4)}{\mcent}{195.701(9)}
\tabc{\sbr\stl^*}{\mcent}{\pone20.7535(5)}{\mcent}{\pone20.7532(7)}{\mcent}{\pone20.753(1)}
\hline
 \end{tabular}\label{xsec:GW-}
  \end{footnotesize}
%
\subsection{\boldmath$gg$ processes}

  \begin{footnotesize}
  \begin{tabular}{|c||ll|ll|ll|}\hline
    \multicolumn{7}{|c|}{$\bf{gg\to X}$} \\\hline
    Final &  \multicolumn{2}{c|}{{\sc madgraph}/{\sc helas}} & 
    \multicolumn{2}{c|}{{\sc o'mega}/{\sc whizard}} & \multicolumn{2}{c|}{{\sc amegic++/sherpa}}\\
    state
    & \multicolumn{1}{c}{0.5 TeV} 
    & \multicolumn{1}{c|}{2 TeV} 
    & \multicolumn{1}{c}{0.5 TeV} 
    & \multicolumn{1}{c|}{2 TeV} 
    & \multicolumn{1}{c}{0.5 TeV} 
    & \multicolumn{1}{c|}{2 TeV} 
    \\\hline\hline 
\tabc{\glo\glo}{\mcent}{13575(2)}{\mcent}{13575.6(1)}{\mcent}{13575.8(7)}
\hline
\tabc{\sul\sul^*}{\mcent}{\phun185.60(2)}{\mcent}{\phun185.615(3)}{\mcent}{\phun185.61(1)}
\tabc{\sur\sur^*}{\mcent}{\phun191.58(2)}{\mcent}{\phun191.590(3)}{\mcent}{\phun191.59(1)}
\tabc{\scl\scl^*}{\mcent}{\phun185.60(2)}{\mcent}{\phun185.612(3)}{\mcent}{\phun185.61(1)}
\tabc{\scr\scr^*}{\mcent}{\phun191.58(2)}{\mcent}{\phun191.588(3)}{\mcent}{\phun191.59(1)}
\tabc{\stl\stl^*}{\mcent}{\phun250.70(2)}{\mcent}{\phun250.71(1)}{\mcent}{\phun250.70(1)}
\tabc{\str\str^*}{\mcent}{\phun180.54(2)}{\mcent}{\phun180.541(3)}{\mcent}{\phun180.54(1)}
\hline
\tabc{\sdl\sdl^*}{\mcent}{\phun184.07(2)}{\mcent}{\phun184.081(3)}{\mcent}{\phun184.09(1)}
\tabc{\sdr\sdr^*}{\mcent}{\phun191.87(2)}{\mcent}{\phun191.875(3)}{\mcent}{\phun191.87(1)}
\tabc{\ssl\ssl^*}{\mcent}{\phun184.07(2)}{\mcent}{\phun184.079(3)}{\mcent}{\phun184.08(1)}
\tabc{\ssr\ssr^*}{\mcent}{\phun191.87(2)}{\mcent}{\phun191.873(3)}{\mcent}{\phun191.86(1)}
\tabc{\sbl\sbl^*}{\mcent}{\phun201.88(2)}{\mcent}{\phun201.884(4)}{\mcent}{\phun201.90(1)}
\tabc{\sbr\sbr^*}{\mcent}{\phun192.52(2)}{\mcent}{\phun192.516(3)}{\mcent}{\phun192.53(1)}
\hline
 \end{tabular}\label{xsec:GG}
  \end{footnotesize}

\clearpage
\subsection{\boldmath$u\bar{u}$ processes}

  \begin{footnotesize}
  \begin{tabular}{|c||ll|ll|ll|}\hline
    \multicolumn{7}{|c|}{$\bf{u\bar{u}\to X}$} \\\hline
    Final &  \multicolumn{2}{c|}{{\sc madgraph}/{\sc helas}} & 
    \multicolumn{2}{c|}{{\sc o'mega}/{\sc whizard}} & \multicolumn{2}{c|}{{\sc amegic++/sherpa}}\\
    state
    & \multicolumn{1}{c}{0.5 TeV} 
    & \multicolumn{1}{c|}{2 TeV} 
    & \multicolumn{1}{c}{0.5 TeV} 
    & \multicolumn{1}{c|}{2 TeV} 
    & \multicolumn{1}{c}{0.5 TeV} 
    & \multicolumn{1}{c|}{2 TeV} 
    \\
\hline
 \end{tabular}\label{xsec:uu}
  \end{footnotesize}

\clearpage
\subsection{\boldmath$d\bar{d}$ processes}

  \begin{footnotesize}
  \begin{tabular}{|c||ll|ll|ll|}\hline
    \multicolumn{7}{|c|}{$\bf{d\bar{d}\to X}$} \\\hline
    Final &  \multicolumn{2}{c|}{{\sc madgraph}/{\sc helas}} & 
    \multicolumn{2}{c|}{{\sc o'mega}/{\sc whizard}} & \multicolumn{2}{c|}{{\sc amegic++/sherpa}}\\
    state
    & \multicolumn{1}{c}{0.5 TeV} 
    & \multicolumn{1}{c|}{2 TeV} 
    & \multicolumn{1}{c}{0.5 TeV} 
    & \multicolumn{1}{c|}{2 TeV} 
    & \multicolumn{1}{c}{0.5 TeV} 
    & \multicolumn{1}{c|}{2 TeV} 
    \\
\hline
 \end{tabular}\label{xsec:dd}
  \end{footnotesize}

\clearpage
\subsection{\boldmath$b\bar{b}$ processes}

  \begin{footnotesize}
  \begin{tabular}{|c||ll|ll|ll|}\hline
    \multicolumn{7}{|c|}{$\bf{b\bar{b}\to X}$} \\\hline
    Final &  \multicolumn{2}{c|}{{\sc madgraph}/{\sc helas}} & 
    \multicolumn{2}{c|}{{\sc o'mega}/{\sc whizard}} & \multicolumn{2}{c|}{{\sc amegic++/sherpa}}\\
    state
    & \multicolumn{1}{c}{0.5 TeV} 
    & \multicolumn{1}{c|}{2 TeV} 
    & \multicolumn{1}{c}{0.5 TeV} 
    & \multicolumn{1}{c|}{2 TeV} 
    & \multicolumn{1}{c}{0.5 TeV} 
    & \multicolumn{1}{c|}{2 TeV} 
    \\
\hline
 \end{tabular}\label{xsec:bb}
  \end{footnotesize}

\clearpage
\subsection{\boldmath$qg$ processes}

  \begin{footnotesize}
  \begin{tabular}{|c||ll|ll|ll|}\hline
    \multicolumn{7}{|c|}{$\bf{qg\to X}$} \\\hline
    Process &  \multicolumn{2}{c|}{{\sc madgraph}/{\sc helas}} & 
    \multicolumn{2}{c|}{{\sc o'mega}/{\sc whizard}} & \multicolumn{2}{c|}{{\sc amegic++/sherpa}}\\
   
    & \multicolumn{1}{c}{0.5 TeV} 
    & \multicolumn{1}{c|}{2 TeV} 
    & \multicolumn{1}{c}{0.5 TeV} 
    & \multicolumn{1}{c|}{2 TeV} 
    & \multicolumn{1}{c}{0.5 TeV} 
    & \multicolumn{1}{c|}{2 TeV} 
    \\\hline\hline 
\tabc{ug\to\sul\glo}{\mcent}{3405.0(5)}{\mcent}{3405.2(3)}{\mcent}{3404.8(2)}
\tabc{ug\to\sur\glo}{\mcent}{3460.0(5)}{\mcent}{3460.0(3)}{\mcent}{3460.4(2)}
\tabc{dg\to\sdl\glo}{\mcent}{3390.0(5)}{\mcent}{3390.5(3)}{\mcent}{3390.0(2)}
\tabc{dg\to\sdr\glo}{\mcent}{3462.5(5)}{\mcent}{3462.5(3)}{\mcent}{3462.0(2)}
\hline
 \end{tabular}\label{xsec:qG}
  \end{footnotesize}

\subsection{Two identical fermions as initial state}

  \begin{footnotesize}
  \begin{tabular}{|c||ll|ll|ll|}\hline
    \multicolumn{7}{|c|}{$\bf{ff\to X}$} \\\hline
    Process &  \multicolumn{2}{c|}{{\sc madgraph}/{\sc helas}} & 
    \multicolumn{2}{c|}{{\sc o'mega}/{\sc whizard}} & \multicolumn{2}{c|}{{\sc amegic++/sherpa}}\\
    & \multicolumn{1}{c}{0.5 TeV} 
    & \multicolumn{1}{c|}{2 TeV} 
    & \multicolumn{1}{c}{0.5 TeV} 
    & \multicolumn{1}{c|}{2 TeV} 
    & \multicolumn{1}{c}{0.5 TeV} 
    & \multicolumn{1}{c|}{2 TeV} 
    \\
\hline
 \end{tabular}\label{xsec:ff-iden}
  \end{footnotesize}

\clearpage
\section{Input Parameters for the LHC and ILC simulations}
\label{sec:bpoint}

\begin{quote}
  \footnotesize
  \begin{verbatim}
BLOCK DCINFO  # Decay Program information
     1   SDECAY      # decay calculator
     2   1.1a        # version number
#
BLOCK SPINFO  # Spectrum calculator information
     1   SOFTSUSY    # spectrum calculator                 
     2   1.9         # version number                      
#
BLOCK MODSEL  # Model selection
     1     0   extSugra                                          
#
BLOCK SMINPUTS  # Standard Model inputs
         1     1.27908957E+02   # alpha_em^-1(M_Z)^MSbar
         2     1.16637000E-05   # G_F [GeV^-2]
         3     1.18700000E-01   # alpha_S(M_Z)^MSbar
         4     9.11876000E+01   # M_Z pole mass
         5     2.50000000E+00   # mb(mb)^MSbar
         6     1.70000000E+02   # mt pole mass
         7     1.77699000E+00   # mtau pole mass
#
BLOCK MINPAR  # Input parameters - minimal models
         3     2.00000000E+01   # tanb                
#
BLOCK EXTPAR  # Input parameters - non-minimal models
        34     5.68797374E+01   # meR(MX)             
        35     1.89750900E+02   # mmuR(MX)            
        36     8.00000000E+02   # mtauR(MX)           
        45    -5.16238332E+02   # mcR(MX)             
#
BLOCK MASS  # Mass Spectrum
# PDG code           mass       particle
        24     7.98256000E+01   # W+
        25     1.14451412E+02   # h
        35     3.00156029E+02   # H
        36     2.99997325E+02   # A
        37     3.10961504E+02   # H+
         5     2.50000000E+00   # b [running mass parameter]
   1000001     4.41227652E+02   # ~d_L
   2000001     4.37876121E+02   # ~d_R
   1000002     4.33747239E+02   # ~u_L
   2000002     4.35113863E+02   # ~u_R
   1000003     4.41227652E+02   # ~s_L
   2000003     4.37876121E+02   # ~s_R
   1000004     4.33747239E+02   # ~c_L
   2000004     4.35113863E+02   # ~c_R
   1000005     2.95364891E+02   # ~b_1
   2000005     3.99917523E+02   # ~b_2
   1000006     4.13841488E+02   # ~t_1
   2000006     9.78880993E+02   # ~t_2
   1000011     2.05024705E+02   # ~e_L
   2000011     2.05651082E+02   # ~e_R
   1000012     1.89267532E+02   # ~nu_eL
   1000013     2.05024705E+02   # ~mu_L
   2000013     2.05651082E+02   # ~mu_R
   1000014     1.89267532E+02   # ~nu_muL
   1000015     1.93593658E+02   # ~tau_1
   2000015     2.16389302E+02   # ~tau_2
   1000016     1.89240110E+02   # ~nu_tauL
   1000021     8.00886030E+02   # ~g
   1000022     4.68440180E+01   # ~chi_10
   1000023     1.12408563E+02   # ~chi_20
   1000025    -1.48090300E+02   # ~chi_30
   1000035     2.36766770E+02   # ~chi_40
   1000024     1.06599344E+02   # ~chi_1+
   1000037     2.37250120E+02   # ~chi_2+
#
BLOCK NMIX  # Neutralino Mixing Matrix
  1  1     8.95603865E-01   # N_11
  1  2    -9.72020087E-02   # N_12
  1  3     4.04193897E-01   # N_13
  1  4    -1.58343869E-01   # N_14
  2  1    -4.03047040E-01   # N_21
  2  2    -5.13608598E-01   # N_22
  2  3     5.77552867E-01   # N_23
  2  4    -4.90093846E-01   # N_24
  3  1    -1.49313892E-01   # N_31
  3  2     1.60265318E-01   # N_32
  3  3     6.53298812E-01   # N_33
  3  4     7.24721361E-01   # N_34
  4  1    -1.14682879E-01   # N_41
  4  2     8.37301025E-01   # N_42
  4  3     2.76153292E-01   # N_43
  4  4    -4.57727201E-01   # N_44
#
BLOCK UMIX  # Chargino Mixing Matrix U
  1  1    -3.90666525E-01   # U_11
  1  2     9.20532273E-01   # U_12
  2  1    -9.20532273E-01   # U_21
  2  2    -3.90666525E-01   # U_22
#
BLOCK VMIX  # Chargino Mixing Matrix V
  1  1    -6.55146178E-01   # V_11
  1  2     7.55502141E-01   # V_12
  2  1    -7.55502141E-01   # V_21
  2  2    -6.55146178E-01   # V_22
#
BLOCK STOPMIX  # Stop Mixing Matrix
  1  1     9.92937358E-01   # cos(theta_t)
  1  2     1.18639802E-01   # sin(theta_t)
  2  1    -1.18639802E-01   # -sin(theta_t)
  2  2     9.92937358E-01   # cos(theta_t)
#
BLOCK SBOTMIX  # Sbottom Mixing Matrix
  1  1     9.13760750E-02   # cos(theta_b)
  1  2     9.95816455E-01   # sin(theta_b)
  2  1    -9.95816455E-01   # -sin(theta_b)
  2  2     9.13760750E-02   # cos(theta_b)
#
BLOCK STAUMIX  # Stau Mixing Matrix
  1  1     7.16384593E-01   # cos(theta_tau)
  1  2     6.97705608E-01   # sin(theta_tau)
  2  1    -6.97705608E-01   # -sin(theta_tau)
  2  2     7.16384593E-01   # cos(theta_tau)
#
BLOCK ALPHA  # Higgs mixing
          -6.49713878E-02   # Mixing angle in the neutral Higgs boson sector
#
BLOCK HMIX Q=  6.12412338E+02  # DRbar Higgs Parameters
         1     1.33393949E+02   # mu(Q)MSSM           
         2     1.94594998E+01   # tan                 
         3     2.43561981E+02   # higgs               
         4     1.06061486E+05   # mA^2(Q)MSSM         
#
BLOCK GAUGE Q=  6.12412338E+02  # The gauge couplings
     1     3.61902434E-01   # gprime(Q) DRbar
     2     6.48956611E-01   # g(Q) DRbar
     3     1.09052463E+00   # g3(Q) DRbar
#
BLOCK AU Q=  6.12412338E+02  # The trilinear couplings
  1  1     0.00000000E+00   # A_u(Q) DRbar
  2  2     0.00000000E+00   # A_c(Q) DRbar
  3  3    -5.80795469E+02   # A_t(Q) DRbar
#
BLOCK AD Q=  6.12412338E+02  # The trilinear couplings
  1  1     0.00000000E+00   # A_d(Q) DRbar
  2  2     0.00000000E+00   # A_s(Q) DRbar
  3  3    -1.84431338E+02   # A_b(Q) DRbar
#
BLOCK AE Q=  6.12412338E+02  # The trilinear couplings
  1  1     0.00000000E+00   # A_e(Q) DRbar
  2  2     0.00000000E+00   # A_mu(Q) DRbar
  3  3    -3.54951850E-01   # A_tau(Q) DRbar
#
BLOCK Yu Q=  6.12412338E+02  # The Yukawa couplings
  1  1     0.00000000E+00   # y_u(Q) DRbar
  2  2     0.00000000E+00   # y_c(Q) DRbar
  3  3     8.97145644E-01   # y_t(Q) DRbar
#
BLOCK Yd Q=  6.12412338E+02  # The Yukawa couplings
  1  1     0.00000000E+00   # y_d(Q) DRbar
  2  2     0.00000000E+00   # y_s(Q) DRbar
  3  3     2.73916822E-01   # y_b(Q) DRbar
#
BLOCK Ye Q=  6.12412338E+02  # The Yukawa couplings
  1  1     0.00000000E+00   # y_e(Q) DRbar
  2  2     0.00000000E+00   # y_mu(Q) DRbar
  3  3     2.03572357E-01   # y_tau(Q) DRbar
#
BLOCK MSOFT Q=  6.12412338E+02  # The soft SUSY breaking masses at the scale Q
         1     5.67130636E+01   # M_1(Q)              
         2     1.89501347E+02   # M_2(Q)              
         3     8.04258574E+02   # M_3(Q)              
        21     7.62419102E+04   # mH1^2(Q)            
        22    -2.17208514E+04   # mH2^2(Q)            
        31     1.99736710E+02   # meL(Q)              
        32     1.99736710E+02   # mmuL(Q)             
        33     1.99710725E+02   # mtauL(Q)            
        34     5.68797374E+01   # meR(MX)             
        35     1.89750900E+02   # mmuR(MX)            
        36     8.00000000E+02   # mtauR(MX)           
        41     4.08231245E+02   # mqL1(Q)             
        42     4.08231245E+02   # mqL2(Q)             
        43     3.73390800E+02   # mqL3(Q)             
        44     4.07778323E+02   # muR(Q)              
        45    -5.16238332E+02   # mcR(MX)             
        46     9.51532103E+02   # mtR(Q)              
        47     4.08215309E+02   # mdR(Q)              
        48     4.08215309E+02   # msR(Q)              
        49     2.58317593E+02   # mbR(Q)              
  \end{verbatim}
\end{quote}

\begin{table}[hb!]
\begin{gather*}
  \begin{array}{|c|l||c|l|}
    \hline
    \text{Particle} & \Gamma\ [\GeV] & \text{Particle} & \Gamma\ [\GeV]
    \\
    \hline
    Z               & 2.4148 &
    \neub           & 5.1100\times 10^{-5}     
    \\
    h               & 5.0080\times 10^{-3} &
    \neuc           & 1.1622\times 10^{-2} 
    \\
    H               & 2.2924 &
    \neud           & 1.0947 
    \\
    A               & 2.7750 &
    \sbl            & 0.53952 
    \\
    & &
    \sbr            & 3.4956
    \\ \hline
    \end{array}
\end{gather*}
\caption{Relevant tree-level particle widths using the input of 
Appendix~\ref{sec:bpoint}.}
\label{tab:decay}
\end{table}


\end{document}